\documentclass[useAMS,usegraphicx,usedcolumn,usenatbib,psfig]{mn2e}
\usepackage{psfig}
\usepackage{natbib}
\usepackage{txfonts}
\usepackage{natbib}
\usepackage{graphicx}
\usepackage{epsfig}
\usepackage{graphics}
\usepackage{color}
\usepackage{subfigure}
\usepackage{amsfonts}
\usepackage{amssymb}

\def\prd{Phys.~Rev.~D} % PRD
\def\prl{Phys.~Rev.~Lett.} % PRL
\def\apj{ApJ} % Astrophysical Journal
\def\apjs{ApJS} % Astrophysical Journal, Supplement
\def\mnras{MNRAS} % Monthly Notices of the RAS
\def\aap{A\&A} % Astronomy and Astrophysics
\def\aaps{A\&AS} % Astronomy and Astrophysics
\def\nat{Nat} % Nature

\def\be#1{\begin{equation}\label{#1}}
\def\ee{\end{equation}}

\input epsf

\title[Blind component separation for polarized observations of the
CMB]{Blind component separation for polarized
  observations of the CMB}

\author[J.~Aumont, J.~F.~Mac{\'\i}as-P\'erez]{
  J.~Aumont, J.~F.~Mac{\'\i}as-P\'erez\\
  Laboratoire de Physique Subatomique et de Cosmologie, 53 Avenue des
  Martyrs, 38026 Grenoble Cedex, France}

\begin{document}

\maketitle

\date{\today}

\begin{abstract}
We present in this paper the {\sc PolEMICA} (Polarized
Expectation-Maximization Independent Component Analysis) algorithm which is an extension to polarization of
the SMICA (Spectral Matching Independent Component Analysis)
temperature multi-detectors multi-components (MD-MC) component separation method \cite{delabrouille}.
This algorithm allows us to estimate blindly in harmonic space multiple physical 
components from multi-detectors
polarized sky maps. Assuming a linear noisy mixture of components we
are able to reconstruct jointly
the anisotropies electromagnetic spectra of the components for each mode $T$, $E$ and $B$, as well as
the temperature and polarization spatial power spectra, $TT$, $EE$, $BB$, $TE$, $TB$ and $EB$ for 
each of the physical components and for the noise on each of the detectors.
{\sc PolEMICA} is specially developed to estimate the CMB temperature and polarization
power spectra from sky observations including both CMB and foreground emissions.
This has been tested intensively using as a first approach full sky simulations of the Planck satellite polarized channels for
a 14-months nominal mission assuming a \emph{simplified} linear sky model including CMB, and optionally
Galactic synchrotron emission and a \emph{Gaussian} dust emission. Finally, we have applied
  our algorithm to more \emph{realistic} Planck full sky simulations,
  including synchrotron, \emph{realistic} dust and free-free emissions.
%From this analysis we conclude that
%although the foreground contribution can be removed, it reduces significantly the
%precision to which the CMB polarization can be measured.
\end{abstract}

\begin{keywords}
-- Cosmic microwave background -- Cosmology: observations -- Methods:
data analysis
\end{keywords}

%________________________________________________________________

%+++++++++++++++++++++++++++++++++++++++++++++++++++++++++++++++++++++++++++++
%+++++++++++++++++++++++++++++++++++++++++++++++++++++++++++++++++++++++++++++
%+++++++++++++++++++++++++++++++++++++++++++++++++++++++++++++++++++++++++++++
%+++++++++++++++++++++++++++++++++++++++++++++++++++++++++++++++++++++++++++++
%+++++++++++++++++++++++++++++++++++++++++++++++++++++++++++++++++++++++++++++

\section[]{Introduction}
\label{intro}

Mapping the Cosmic Microwave Background (CMB) polarization is one
of the major challenges of future missions of observational cosmology.
CMB polarization is linear and therefore can be described by the first three Stokes
parameters I, Q and U which are generally combined to produce three fields
(modes), $T$, $E$ and $B$ \cite{zaldarriaga}.
The polarization of the CMB photons carries extra physical informations that are not
accessible by the study of the temperature anisotropies. Therefore its measurement helps breaking down the degeneracies on cosmological
parameters as encounter with temperature anisotropies measurements only \cite{zalda_spergel}. 
Furthermore, the study of the CMB polarization is also a fundamental tool to estimate the
energy scale
of inflation which has  been proposed to solve the problems of flatness, of isotropy and of the
seed perturbations for the formation of the structures in the Universe. 
Inflationary models predict the presence of tensor
perturbations of the metric which will lead to an unique
signature in the CMB polarization $B$ modes. The detection of the latter would be a strong
proof of such an epoch and also a way to constrain the energy scale at
which inflation occurs by measuring the tensor to scalar ratio, $r$ \cite{turner}.

Since the beginning of the CMB anisotropies observations with the COsmic Background Explorer
(COBE) \cite{cobe}, a great amount of experiments have been designed to determine
the CMB temperature angular power spectrum \cite{saskatoon, toco, boom1,
maxima1, maxima2, boom2, dasi, cbi, vsa, archeops_cl, wmap_cl,
barkats, cbi_04, dasi_04, archeops_cl2, boom3}.
By contrast, the polarization anisotropies, which are
between 2 and 5 orders of magnitude weaker than temperature
ones are not accurately measured yet. A first
detection of the CMB $E$ modes has been performed by DASI \cite{dasi_pol,dasi_04},
 CAPMAP \cite{barkats}, CBI \cite{cbi_04} and more recently by {\sc BOOMERanG}
\cite{boomee} and WMAP \cite{wmap_te}. The $TE$ temperature-polarization cross correlation has
been  measured by WMAP \cite{wmap_te} and {\sc BOOMERanG} \cite{boomte}. No detection 
of the CMB $B$ modes has been reported yet. Nevertheless,
  constraints on the tensor to scalar ratio, $r$, have been set by
  the WMAP team. They set an upper limit of $r<0.55$ (95\% CL)
  \cite{wmap_cosmo06} for the temperature and polarization analysis and of $r<2.2$ (95$\%$ CL)
  \cite{wmap_te} for a polarization-only analysis.

The detection of such low signals is possible by improving the instrumental
sensitivity, but this is not the only issue in the determination of
the CMB polarization power spectra. Other astrophysical emissions 
as for example the diffuse Galactic emission including free-free, dust and synchrotron
and the extragalactic-sources emissions also
contribute to the sky brightness at the frequencies of interest for CMB studies, 
and therefore must be efficiently subtracted. These \emph{foregrounds} are
particularly important for the study of the CMB polarization. Excluding
the free-free emission which is not polarized, the other contributions
are expected to be significantly polarized with similar power on the $E$ and $B$ modes. 
Recent measurements of the Galactic synchrotron polarization emission
at 1.41~GHz \cite{wolleben} and at 23 GHz by WMAP \cite{wmap_te}
show this emission is significantly polarized at large angular scales. 
Further, Archeops measurements at 353~GHz show that the Galactic dust
diffuse emission is polarized up to a level of 5 to 10 \% both in the
Galactic center \cite{benoit_polar} and at high Galactic
latitudes \cite{tearcheops}. Finally, for the polarization of
extragalactic point sources the sparsity of the data available makes
reliable predictions difficult \cite{tucci04,hildebrand96}.

A direct subtraction of these foreground contributions on the CMB data will require
an accurate knowledge of their spatial distributions and of the electromagnetic
spectra of their anisotropies. For the synchrotron emission a full sky map at 408 MHz in temperature
is available \cite{haslam} and more recently the WMAP
team provided a map at 30~GHz from the MEM decomposition of the
first year observations \cite{wmapforegrounds}. 
A fake polarized synchrotron emission template was constructed by \cite{giardino}
based on the Parkes 2400 MHz \cite{parkes} and Haslam 408 MHz
\cite{haslam} surveys. 
Furthermore, the electromagnetic spectrum of synchrotron anisotropies and 
its spatial distribution are neither accurately known in temperature
nor polarization although a first estimate was produced by
\cite{giardino}. Recently, the 23 GHz polarized WMAP data is used
as a tracer of the synchrotron polarization \cite{wmap_te}. 
For the thermal dust emission a full sky map at 100~$\mu$m as well as templates 
for CMB use were extracted from the IRAS and FIRAS data \cite{sfd,fds}. 
No realistic template exists for the dust polarized emission, although
a fake one, based on the polarization angles measured at 23 GHz has
been constructed by the Planck collaboration\footnote{Planck Sky
  Model, {\tt http://www.cesr.fr/$\sim$bernard/PSM/}}. 
The dust emission in temperature can be approximated by a
grey body spectrum of mean temperature 17~K and emissivity between 1.7 and 2.2 \cite{fds,lagache03}.
Currently no measurement on the electromagnetic spectrum of the dust polarized emission is
available although it is expected to be the same that for temperature \cite{kelbe}.

To try to overcome the above limitations, a great amount of work has been dedicated to design and implement algorithms
for component separation which can discriminate between CMB and foregrounds. These methods can also
extract, directly from the CMB data, the emission properties of foregrounds. Wiener filtering has
been successfully tested assuming known Gaussian priors for each component
and with the electromagnetic spectrum of the anisotropies as an input \cite{tegmark,bouchet}. Maximum entropy 
based methods (MEM), assuming entropic priors for the spatial distribution of
each of the component, have been intensively used for small sky patches \cite{hobson} and
extended to full sky analysis \cite{stolyarov_02}. They
were adapted to account for spatial anisotropies in the electromagnetic
spectra \cite{stolyarov_04}. More recently, \cite{eriksen} has developed a new
method to perform CMB component separation by parameter estimation and applied
it to temperature simulations of the Planck satellite experiments.  
Independent Component Analysis (ICA) techniques have also been applied to Planck
simulations in temperature \cite{maino} and extended to polarization \cite{baccigalupi,stivoli} using the {\sc FastICA}
algorithm. These methods require no prior on the spectral or spatial distribution of the
components but can not make use of the available physical knowledge on the foreground and CMB emissions. 
In addition, the Spectral Matching Independent Component Analysis  (SMICA) \cite{delabrouille} 
has been developed to consider both the fully blind analysis for which no prior is assumed and
the semi-blind analysis incorporating previous physical knowledge on the astrophysical components.
This algorithm, based
on the Expectation-Maximization algorithm (EM) \cite{em}, uses the spectral
diversity of the components and was developed for temperature only. 
We present in this paper, {\sc PolEMICA} (Polarized
Expectation-Maximization Independent Component Analysis), an extension of this method to polarization
including both the blind and semi-blind analysis.

This paper is organized as follows. A simple model of the microwave sky emission 
in temperature and polarization is described in section~\ref{simus}. Section~\ref{mdmc} presents
the {\sc PolEMICA} multi-detectors multi-components (MD-MC) blind component separation algorithm.  Section~\ref{simus_list} describe
the simulations of the Planck satellite experiment used for testing
the algorithm. 
We present in section \ref{testing} the application of {\sc PolEMICA}
to the Planck simulations with a \emph{simplified} model to test the
algorithm's performances. Finally, in section
  \ref{realistic_model}, we apply our algorithm to more
  \emph{realistic} Planck simulations and discuss the separability
  problem in this case. We summarize and conclude in section~\ref{conclusions}.

% info de la polar CMB

% ordres de grandeur

% probleme domination foregrounds

% separation de composantes

% plan

%+++++++++++++++++++++++++++++++++++++++++++++++++++++++++++++++++++++++++++++
%+++++++++++++++++++++++++++++++++++++++++++++++++++++++++++++++++++++++++++++
%+++++++++++++++++++++++++++++++++++++++++++++++++++++++++++++++++++++++++++++
%+++++++++++++++++++++++++++++++++++++++++++++++++++++++++++++++++++++++++++++
%+++++++++++++++++++++++++++++++++++++++++++++++++++++++++++++++++++++++++++++

\section[]{Model of the microwave and sub-mm sky}\label{simus}

\subsection{Multi-detectors Multi-components model}
\indent To constrain cosmological models, CMB experiments have to
reach an accuracy which is well below the expected level of
contamination from astrophysical foregrounds, in temperature
and even more critically in polarization. Therefore, 
an efficient separation between CMB and foregrounds is crucial for the success
of future polarization experiments. To perform such a
separation, the diversity of the electromagnetic
spectra of the anisotropies and of the spatial spectra of the components is generally used. 
Observations from a multi-band instrument can
be modeled as a linear combination of multiple physical components
leading to what is called a Multi-Detectors Multi-Components (MD-MC) modeling.

% +++++++++++++++++++++++++++++++++++++++++++++++++++++++++++++++++++++
% + Description of the different methods, here or in Introduction
% part.
% +++++++++++++++++++++++++++++++++++++++++++++++++++++++++++++++++++++

\indent  Assuming an experiment with $n_\nu$
detector-bands at frequencies $\nu_i$ and $n_c$ physical components
in the data, for each Stokes parameter ($I$,
$Q$ and $U$) and for each pixel on the sky map we can write

\be{datareal}
y^{\nu}({\bf r})=A^{\nu}_c\cdot s^{c}({\bf r})+n^{\nu}({\bf r})
\ee

\noindent where $s^{c}({\bf r})$ is the map of the $c$
component, $n^{\nu}({\bf r})$ refers to the noise map for each $\nu_i$
band and $A^{\nu}_c$ which is called the \emph{mixing matrix}, gives the electromagnetic spectrum behavior for the component $c$
and frequency  $\nu_i$. Beam smoothing and filtering effects are not considered in
this work. 

As in the temperature case, it is more convenient to work in spherical
harmonics space, where equation
(\ref{datareal}) can be rewritten  independently for each $\{\ell,m\}$ assuming a full
sky coverage. Thus equation (\ref{datareal}) reads for $X=\{T,E,B\}$ and
for each frequency band and for each $\{\ell,m\}$

\be{data}
y^{\nu,X}_{\ell m}=\sum_{c=1}^{n_c}A^{\nu,X}_cs^{c,X}_{\ell
  m}+n^{\nu,X}_{\ell m}
\ee

\noindent where $y^{\nu,X}_{\ell m}$ is a vector of size $(3\cdot
n_\nu\cdot n_\ell\cdot n_m)$, $s^{c,X}_{\ell m}$ is a $(3\cdot
n_c\cdot n_\ell\cdot n_m)$ vector and $n^{\nu,X}_{\ell m}$ is a vector
of the same size than $y^{\nu,X}_{\ell m}$. $A^{\nu,X}_c$
is a matrix of $(3\cdot n_\nu)\times(3\cdot n_c)$ elements formed from the mixing
matrix of each of the modes, $T$, $E$ and $B$.

The aim of the component separation algorithm presented in this paper
is to extract $A^{\nu,X}_c$, $s^{c,X}_{\ell m}$ and $n^{\nu,X}_{\ell m}$ 
from the $y^{\nu,X}_{\ell m}$ sky observations.

\subsection{Simulated microwave and sub-mm sky}

Following the MD-MC model discussed above and given an observational setup, 
we construct, using the HEALPix pixelization
scheme \cite{healpix} and in CMB temperature
units, fake $I$, $Q$ and $U$ maps of the sky at each of the instrumental frequency bands.
For these maps we consider three main physical components
in the sky emission: CMB, thermal dust and synchrotron. Concerning emissions that are not expected to be significantly
  polarized, we have optionally considered unpolarized free-free emission and not
taken into account the SZ emission as we are interested in diffuse emissions. When no free-free is considered
and in any cases for the SZ emission, we suppose that they have been successfully
removed from our sky maps in temperature. We have assumed that the
free-free emission is not polarized. Instrumental noise
is modeled as white noise.

\subsubsection*{CMB}
 
  The CMB component map is randomly generated from the
  polarized CMB angular power spectra computed with the CAMB software \cite{camb}
  for a set of given cosmological parameters. In the following we have used $H_0=71\ {\rm
    km}\cdot{\rm s}^{-1}\cdot{\rm Mpc}^{-1}$, $\Omega_b=0.044$,
  $\Omega_m=0.27$, $\Omega_\Lambda=0.73$ and $\tau=0.17$ that are the
  values of the cosmological concordance model according to the WMAP
  one year results
  \cite{wmappc}. We also consider gravitational lensing effects
   as described in \cite{challinor,weaklensinghu,weaklensingchallinor,weaklensingokamoto}.

\subsubsection*{Synchrotron} 

  For the diffuse Galactic synchrotron emission we use the 
  template maps in temperature and in polarization provided by \cite{giardino}.
  These template maps were derived in temperature directly from the Haslam map at 408 MHz
  \cite{haslam}. The polarization maps in $Q$ and $U$ were constructed from the
  intensity map from a constrained realization of the polarization angles using
  the Parkes 2400~MHz survey \cite{parkes}. A template of the spatial variations of the synchrotron
  spectral index is also provided by \cite{giardino}.
  Here we have chosen to use a constant spectral index equal to the mean of the spectral index map, $\alpha=-2.77$,
  so that the simple linear model of the data holds. A more realistic
  treatment of the synchrotron emission will require a specific model of the spectral
  index spatial variations in order to ensure separability between components.
        
 % Thus, the electromagnetic spectrum of synchrotron in our simulations is given by 
 % a power law with constant index $\alpha=-2.77$. 

\subsubsection*{Dust} 

%  In the case of the thermal dust we dispose of few 
%  observational data of the polarized diffuse emission and to date no template 
%  for this is available. %In temperature, we could have used
%    template maps from \cite{sfd,fds} but in this case, we encounter
%    problems for separating {\it blindly} this dust component from the synchrotron
%    one because of their strong spatial correlation. For a short
%    description of this temperature separability problems, see
%    \cite{maino} and \cite{patanchon}.
 We have used through this
    article two different dust models.

    $\star$ Simplified-\emph{dust model}: We have first considered a simplified dust model which is Gaussian
    and
    derived from the power-law model from \cite{prunet}
  to describe the dust angular power spectra in temperature and in
  polarization. This model, although not fully realistic, is not
  spatially correlated to the synchrotron emission and helps
  us to extensively test the properties of the separation method. 
  We have renormalized this model to mimic at large angular scales
  the $TE$ cross power spectrum measured by Archeops at 353~GHz \cite{tearcheops}.
  The rms of the final dust map is probably overestimated as we do not account
  for the variation of the dust emission with respect to Galactic latitude.
  The power spectra models are computed at 100 GHz in $\mu {\rm K}_{\rm RJ}$ units. 
  $I,Q$ and $U$ full-sky maps are generated randomly from
  these power spectra. We extrapolate them to each of the frequency of interest  by assuming
  a grey body spectrum with an emissivity of 2. Finally, the $\mu {\rm K}_{\rm RJ}$ maps are converted
  into $\mu {\rm K}_{\rm CMB}$ units. 

  $\star$ Realistic-\emph{dust model}: Secondly we have used the
    \emph{Planck Sky Model}\footnotemark[1] polarized dust
  template \cite{baccigalupi_psm}. It was modeled using model number 7 of \cite{fds}. This
  model is normalized to the IRAS 100 $\mu$m emission map produced by
  \cite{sfd}. For polarization, a constant polarization degree of 5\%
  is assumed, and the same polarization angles than for the
  synchrotron model are used. For both temperature and polarization,
  we assume a grey body emission with an emissivity of 2.

\subsubsection*{Free-free}

The free-free component is derived from the
  \emph{Planck Sky Model}\footnotemark[1]. It is based on the H-$\alpha$ template by
  \cite{free-free}. We have assumed a constant spectral index of
  -2.1 in Rayleigh-Jeans units. Expected to be no significantly
  polarized except in particular HII regions (less than 10\% \cite{free-free_polar}), the free-free
  emission $Q$ and $U$ maps are set to zero.

\footnotetext[1]{{\tt http://www.cesr.fr/$\sim$bernard/PSM/}}

\subsubsection*{Noise}
  Noise maps for each channel are generated from white noise realizations
  normalized to the nominal level of instrumental noise for that channel. \\

\begin{table}
    \begin{tabular}{|c||c|c|c|c|c|c|c|}
      \hline
      {\tiny $\nu\ $(GHz)} & {\tiny 30} & {\tiny 40} & {\tiny 70} &
      {\tiny 100} & {\tiny 143} & {\tiny 217} & {\tiny 353}\\
      \hline
      {\tiny CMB} & {\tiny 1.0} & {\tiny 1.0} & {\tiny 1.0} &
      {\tiny 1.0} & {\tiny 1.0} & {\tiny 1.0} & {\tiny 1.0}\\
      \hline
      {\tiny Sync.} & {\tiny 1.0} & {\tiny 0.46} & {\tiny 0.11} &
      {\tiny 0.045} & {\tiny 0.021} & {\tiny 0.012} & {\tiny 0.014}\\
      \hline
      {\tiny Dust} & {\tiny 0.0006} & {\tiny 0.001} & {\tiny 0.003} &
      {\tiny 0.008} & {\tiny 0.021} & {\tiny 0.088} & {\tiny 1.0}\\
      \hline
      {\tiny Free-free} & {\tiny 1.0} & {\tiny 0.56} & {\tiny 0.19} &
      {\tiny 0.10} & {\tiny 0.061} & {\tiny 0.046} & {\tiny 0.071}\\
%      \hline
%      {\tiny Farad. rot.} & {\tiny 1.0} & {\tiny 0.56} & {\tiny 0.18}
%      & {\tiny 0.09} & {\tiny 0.044} & {\tiny 0.019} & {\tiny
%      0.007}\\
      \hline
      {\tiny Noise $I$} & {\tiny 4.12} & {\tiny 4.03} & {\tiny 4.06} &
      {\tiny 1.47} & {\tiny 1.0} & {\tiny 1.47} & {\tiny 4.54}\\
      \hline
      {\tiny Noise $P$} & {\tiny 2.91} & {\tiny 2.95} & {\tiny 2.98}
      & {\tiny 1.51} & {\tiny 1.0} & {\tiny 1.51} & {\tiny 4.59}\\
      \hline
    \end{tabular}
  \caption{Electromagnetic spectrum of the anisotropies, in arbitrary
    units, of the CMB, dust, synchrotron and free-free components at 
    each of the Planck channels. Relative noise levels in the temperature and
    in the polarization maps taking as reference the 143 GHz channel
    are also displayed.\label{tab:amat_tab}}
\end{table}

The electromagnetic spectra of the anisotropies, in arbitrary units, for the CMB, dust and synchrotron emissions
are displayed in table~\ref{tab:amat_tab} (these are the values used in  the mixing matrix $A$) 
for the Planck satellite simulations
presented in section~\ref{simus_list}. We also present the relative noise level taking as
reference the 143~GHz channel. The noise levels used at 143~GHz are 
6.3 $\mu$K$_{\rm CMB}$ (in temperature) and 12.3 $\mu$K$_{\rm CMB}$ (in polarization)
per square pixels of side 7 arcmin and for a 14-months Planck mission \cite{bluebook}. 

% We stress the fact that our sky model
%  is a simplified one in many ways but especially for the dust
%modeling as discussed above. The aim is to show that the method presented in this paper
%can deal simultaneously with temperature and polarization
%and that when considering fully uncorrelated components
%(main assumption of the algorithm) the separation is possible.
%Introducing a realistic dust model requires a specific treatment  
%of the cross-correlation in temperature between dust and synchrotron.
%This is beyond the scope of this paper.
% VOICI TA PHRASE AU CAS OU
%The aim of this paper is to show the performances of the
%algorithm in this case, neglecting the separability problem in
%temperature between synchrotron and dust and to concentrate on the
%innovation of this method which is to allow to deal jointly with
%temperature and polarization. We will test our method under a more
%realistic dust model in a forthcoming paper.}

%+++++++++++++++++++++++++++++++++++++++++++++++++++++++++++++++++++++++++++++
%+++++++++++++++++++++++++++++++++++++++++++++++++++++++++++++++++++++++++++++
%+++++++++++++++++++++++++++++++++++++++++++++++++++++++++++++++++++++++++++++
%+++++++++++++++++++++++++++++++++++++++++++++++++++++++++++++++++++++++++++++
%+++++++++++++++++++++++++++++++++++++++++++++++++++++++++++++++++++++++++++++

\section[]{A MD-MC component separation method for polarization}\label{mdmc}

\subsection{MD-MC model for the temperature and polarization power spectra.}
To reduce the number of unknown parameters in the model 
described by equation (\ref{data}), it is interesting to 
rewrite this equation in terms of the temperature and polarization auto and
cross power spectra. This will considerably reduce the computing
time with no loss of information. 

We define
the density matrices associated with the data, $y$, the physical components, $s$ and the
noise, $n$, as follows

\begin{eqnarray}
\label{densmat}
\qquad \qquad R_{\chi}(\ell) & \equiv & \langle\chi^{i,X}_{\ell
  m}\cdot\chi^{i',X'\dagger}_{\ell m}\rangle,\ \ \chi=\{y,s,n\} \nonumber\\
\qquad \qquad & = & \frac{1}{2 \ell+1} \sum_{l= -m}^{m} \chi^{i,X}_{\ell
  m}\cdot\chi^{i',X'\dagger}_{\ell m}
\end{eqnarray}

where $i$ represents frequency, $\nu$, for the data and noise matrices and 
component, $c$, for the physical-components matrix.
Averaging over bins on $\ell$ we obtain

\be{densmatbin}
R_{\chi}(b)\equiv\frac{1}{n_b}\sum_{\ell\in\mathcal{D}_b}\langle\chi^{i,X}_{\ell
  m}\cdot\chi^{i',X'\dagger}_{\ell m}\rangle
\ee

\noindent where $\mathcal{D}_b$ is the set of $\ell$ values which
contributes to bin $b$ and $n_b$ is the number of such multipoles.
In the following, $N_{b}$, represents the total number of bins used in the analysis.

For each bin $b$, equation~(\ref{data}) reads

\be{dataspec}
R_y(b) = AR_s(b)A^T+R_n(b)
\ee

\noindent where $R_y(b)$ and $R_n(b)$ are
$(n_\nu\cdot3)\times(n_\nu\cdot3)$ matrices and $R_s(b)$ is a
$(n_c\cdot3)\times(n_c\cdot3)$ matrix. 

To fully understand the component separation algorithm described below
it is interesting to have a closer look to the content of the
three density matrices defined above (see appendix for a
concrete example).

$R_y(b)$ represents the input density matrix computed from the observed
multi-band data. This matrix is composed of  $3\times3$ symmetric blocks
each of them containing in the diagonal the auto-power spectra,
$TT$, $EE$, $BB$ and in the off-diagonal the cross-power spectra $TE$, $EB$ and $TB$. 
A single block represents either the
auto-correlation of a single channel (for diagonal blocks) or the
cross-correlation between two channels (for off-diagonal blocks).

Assuming that the physical components in the data are statistically
independent and uncorrelated makes the $R_s(b)$ 
a $3\cdot n_c\times3\cdot n_c$ block diagonal matrix. As above
each block, corresponding to the $c^{\rm th}$ physical component, contains
in the diagonal the auto-power spectra,
$TT$, $EE$, $BB$ and in the off-diagonal the cross-power spectra $TE$, $EB$ and $TB$. 

We also assume that the noise is uncorrelated between channels 
and therefore, $R_n(b)$ is a diagonal matrix containing the noise
auto power spectra  $TT$, $EE$ and $BB$ for each of the channels.

\subsection{Spectral matching algorithm}

\label{em_algorithm}

From equation~(\ref{dataspec}) we observe that, for each bin $b$, the data density
matrix $R_y(b)$, of size $9\times n_\nu^2$,
is fully defined by the set of parameters $\theta(b)=\{A,R_s(b),{\rm diag}(R_n(b))\}$
which corresponds to a total of $3\times n_\nu \times n_c + 6\times n_c + 3\times n_\nu$ parameters.
This indicates that from the CMB data set and under the hypothesis presented above
it is possible to simultaneously estimate the mixing matrix, the physical component's
temperature and polarization power spectra and the noise's temperature and
polarization power spectra for each of the channels. Further,
assuming white noise in the maps only three noise parameters per channel
(for TT, EE and BB) need to be estimated for the entire range in $\ell$.
This reduces the overall set of parameters to 
$3\times n_\nu \times n_c + 6\times n_c \times N_{b} + 3\times n_\nu$
where $N_{b}$ is the number of bins.\\

\subsubsection*{The likelihood function}
To estimate the above parameters from the data we have extended
to the case of polarized data the spectral matching algorithm
developed by \cite{delabrouille} for temperature only.
The key issue of this method is to estimate these parameters, or some of them  (for a semi-blind analysis),
by finding the best match between the model density matrix, $R_y(b)$, computed for $\theta(b)$
and the data density matrix $\tilde{R}_y(b)$ obtained from
the multi-channel data. Assuming that the different physical components
and the noise are realizations of Gaussian stationary fields (Wittle approximation), the
log-likelihood function of the form
\be{loglike}
\mathcal{-L}=\sum_{b=1}^{N_{b}} n_b\bigg[{\rm
  Tr.}\Big(\tilde{R}_y(b)R^{-1}_y(b)\Big)+\log\det\Big(R_y(b)\Big)\bigg]+{\rm c.}
\ee
is a reasonable measure of the mismatch between data and model. 

%Here, the log-likelihood measures a mean gradient between two
%matrices that are the empirical density matrix and its expectation
%value. Thus we can write the $\Psi(\theta)$ criterion that depends of
%the Kullback divergence \cite{kullback}:

%\be{loglikekull}
%\mathcal{L}=\sum_{q=1}^Qn_qD\Big(\tilde{R}_y(q),R_y(q)\Big)
%\ee

%\noindent where the Kullback divergence reads:

%$$
%D\Big(\tilde{R}_y(q),R_y(q)\Big)\equiv{\rm Tr.}\Big(\tilde{R}_y(q)R^{-1}_y(q)\Big)
%$$
%\be{kull}
%\qquad\qquad-\log\det\Big(\tilde{R}_y(q)R^{-1}_y(q)\Big)+{\rm cst}
%\ee
\subsubsection*{EM algorithm}

The maximization of the likelihood function is achieved via the
Expectation-Maximization algorithm (EM) \cite{em}. This algorithm
will process iteratively from an initial value of the parameters
$\theta_{0}(b)$ following a sequence of parameter updates $\theta_{i}(b)$,
called `EM steps'. During the $i^{\rm th}$ E-step we compute the expectation 
value for the likelihood from
the $(i-1)^{\rm th}$ iteration's parameters. The  $i^{\rm th}$ M-step maximizes the likelihood
(i.e. minimizes the log-likelihood) to compute the $i^{\rm th}$ set
of parameters. By construction each EM step improves the spectral fit by maximizing
the likelihood. For a more detailed review of the spectral matching EM
algorithm used here, see \cite{snoussi} and \cite{delabrouille} and 
appendix~\ref{matrix_appendix} for the formalism used to describe the polarized sky model
and data.

In the MD-MC model presented above there is a scale indetermination on the value of $A$ and
$R_s$ and only the product $AR_sA^T$ is scale invariant. Thus, to ease
the convergence of the algorithm, we renormalize each column of $A$ to unity at each
EM iteration and correct the $R_s$ density matrix accordingly so that the product
$AR_sA^T$ is unchanged.

\subsubsection*{Initialization of the algorithm}
\label{initialization}

To start-up the EM algorithm the parameters of the fit, $\theta$, need to be initialized
to reasonable values to avoid exploring local maxima in the likelihood function.

In the case of the mixing matrix, $A$, we can consider, in a first approximation, 
that the electromagnetic spectrum is the
same in temperature and polarization for each of the components. Therefore,
we can concentrate on guessing the electromagnetic spectrum in the temperature
data where we expect the signal to noise to be larger.
When no physical prior is available this can be obtained 
by using the $n_c$ dominant eigenvectors of the data density matrix for temperature only,
$\tilde{R}^{TT}_y$. Each of them represent the change of power with frequency
for the dominant components in the data. Notice that these components
and the physical ones are not necessarily the same.
On one hand, we can have in the data extra components which have not
been identified as for example residual systematics. On the other hand,
the electromagnetic spectrum of the physical components may present
spatial variations as it is, for example, the case for the Galactic synchrotron diffuse emission.
In the following, we will consider that the data contain only identified physical
components with spatially constant electromagnetic spectrum.
If this is not the case, a careful pre-analysis of the initialization
parameters is needed and this is not discussed in this paper.
%Then assuming that the electromagnetic spectrum of the polarized components
%as the $n_c$ columns of $A$ that will
%correspond to the foregrounds spectra, for $T$, $E$ and $B$. The elements of $A$ that
%correspond to CMB are fixed to unity. This do not constitute real
%estimates of $A$ and $R_n$, but a guess in the good direction.

Assuming the mixing matrix previously initialized, the physical-components density matrix
can be obtained from a noiseless fit to the data as follows
\be{initrs}
R_{s,0}(b)={\rm block diag}\Big(\left(A_{0}^{T}A_{0}\right)^{-1}A_{0}^{T}\tilde{R}_y(b)A_{0}\left(A_{0}A_{0}^{T}\right)^{-1}\Big)
\ee

Finally, from $A_{0}$ and $R_{s,0}(b)$ the noise density matrix
is given by
\be{initrn}
R_{n,0}(b)={\rm diag}\Big(\sum_b\frac{n_b}{n}\left(\tilde{R}_y(b)-A_{0}R_{s,0}A_{0}^{T}\right)\Big)
\ee
where $n = \sum n_{b}$.
Here we implicitly assume that the noise is white and not correlated
between different channels.

%Notice that both for the mixing matrix, $A$, and for the noise density matrix, $R_{n}(q)$ the
%initial values presented above are just initial guesses for starting point of the EM algorithm
%and not estimated values.

%+++++++++++++++++++++++++++++++++++++++++++++++++++++++++++++++++++++++++++++
%+++++++++++++++++++++++++++++++++++++++++++++++++++++++++++++++++++++++++++++
%+++++++++++++++++++++++++++++++++++++++++++++++++++++++++++++++++++++++++++++
%+++++++++++++++++++++++++++++++++++++++++++++++++++++++++++++++++++++++++++++
%+++++++++++++++++++++++++++++++++++++++++++++++++++++++++++++++++++++++++++++

\section[]{Simulated observations}\label{simus_list}

%+++++++++++++++++++++++++++++++++++++++++++++++++++++++++++++++++++++++++++++
%+++++++++++++++++++++++++++++++++++++++++++++++++++++++++++++++++++++++++++++
%+++++++++++++++++++++++++++++++++++++++++++++++++++++++++++++++++++++++++++++
%+++++++++++++++++++++++++++++++++++++++++++++++++++++++++++++++++++++++++++++
%+++++++++++++++++++++++++++++++++++++++++++++++++++++++++++++++++++++++++++++
We have first performed various sets of simulations 
of the expected Planck satellite data to intensively test the algorithm presented above.
For each of those, we performed 300 realizations considering full-sky
maps that can contain instrumental noise, CMB, dust, synchrotron and free-free. For each realization the
CMB and noise are changed while dust, free-free and synchrotron are kept unchanged. 
We simulate maps at the LFI and HFI polarized channels,
30, 40 and 70~GHz for LFI and 100, 143, 217 and 353~GHz for HFI.
These maps are in HEALPix pixelization \cite{healpix} and 
correspond to a 14-month survey.

        \begin{enumerate}
%% Simu 512 CMB+DUST+SYNC
        \item {\bf \tt [planck~a]}: We simulate maps at
          $n_{\rm side}$ = 512 (pixels of area $\sim 47\ {\rm arcmin}^{2}$).
          We include in the simulations CMB emission and gravitational
          lensing. The simulations also contain  
          synchrotron and \emph{simplified} thermal
          dust emissions as described in section
          \ref{simus}. This $n_{\rm side}$ permits the
          reconstruction of the angular power spectra up to $\ell\simeq1500$. 
          The reconstructed spectra will be
          averaged over bins of size 20 in $\ell$. With these simulations we test the separation
          method at small angular scales. The simulations contain
          also instrumental noise.
%% Simu 128 CMB
        \item {\bf \tt [planck~b]}: We simulate maps at $n_{\rm side}$ = 128 
          (pixels of area $\sim 755$ arcmin$^{2}$) for CMB only. 
          This $n_{\rm side}$ permits the reconstruction of the power spectra
          up to $\ell\simeq400$. This maximum $\ell$ value is enough to study
          the effect of gravitational waves in the $BB$ spectrum which
          is maximum around $\ell=100$ for the concordance model. 
          The reconstructed spectra are averaged over bins of size 10 in $\ell$.
          These simulations were performed for fast and intense test of the
          algorithm in the easiest possible case. The simulations contain
          also instrumental noise.
%% Simu 128 CMB + DUST + SYNC
        \item {\bf \tt [planck~c]}: We simulate maps at  $n_{\rm side}$ = 128 
          including CMB, \emph{simplified} thermal dust and synchrotron emissions.
          The reconstructed spectra are averaged over bins of
          size 10 in $\ell$. These simulations were performed to check the
          impact of foregrounds in the reconstruction of the CMB power spectra. The simulations contain
          also instrumental noise.
        \item {\bf \tt [planck~d]}: Same as {\bf \tt [planck~c]}
            simulations except that \emph{simplified} dust has been
            replaced by \emph{realistic} dust and free-free emission
            has been included.
          
        \end{enumerate}

%\begin{figure*}
%\centering
%\plotancho{./cmbalone_r0.7_300sims.ps}
%\caption[fig:figure1]{Reconstruction of the CMB power spectra for
%$C_\ell^{TT}$, $C_\ell^{EE}$, $C_\ell^{BB}$, $C_\ell^{TE}$,
%$C_\ell^{TB}$ and $C_\ell^{EB}$ at 100 GHz in $\mu {\rm K}_{\rm CMB}^2$
%performed on full sky maps for the {\bf \tt [planck~b]} simulations. Crosses
%represent the reconstructed spectra, solid lines the model used to
%make the simulations (see section \ref{simus}), where the scalar to
%tensor ratio $r$ is equal to 0.7. Error bars are the dispersion over
%$\sim$ 150 full sky simulations.}
%\label{fig:planckcmb128}
%\end{figure*}

%+++++++++++++++++++++++++++++++++++++++++++++++++++++++++++++++++++++++++++++
%+++++++++++++++++++++++++++++++++++++++++++++++++++++++++++++++++++++++++++++
%+++++++++++++++++++++++++++++++++++++++++++++++++++++++++++++++++++++++++++++
%+++++++++++++++++++++++++++++++++++++++++++++++++++++++++++++++++++++++++++++
%+++++++++++++++++++++++++++++++++++++++++++++++++++++++++++++++++++++++++++++
%+++++++++++++++++++++++++++++++++++++++++++++++++++++++++++++++++++++++++++++

\section[]{Testing and performances on the \emph{simplified}-model}\label{testing}

%+++++++++++++++++++++++++++++++++++++++++++++++++++++++++++++++++++++++++++++
%+++++++++++++++++++++++++++++++++++++++++++++++++++++++++++++++++++++++++++++
%+++++++++++++++++++++++++++++++++++++++++++++++++++++++++++++++++++++++++++++
%+++++++++++++++++++++++++++++++++++++++++++++++++++++++++++++++++++++++++++++
%+++++++++++++++++++++++++++++++++++++++++++++++++++++++++++++++++++++++++++++
%+++++++++++++++++++++++++++++++++++++++++++++++++++++++++++++++++++++++++++++

%For testing the method, we will simulate some $I$, $Q$ and $U$ full
%sky maps at different frequencies from which
%we will compute the auto and cross power spectra. Then the EM algorithm is performed
%on the density matrix containing all the correlations from this set of
%maps. The EM algorithm can be performed with different levels of
%freeness:
We have applied the MD-MC {\sc PolEMICA} component separation algorithm
to the \emph{simplified} simulations presented above. For each set of simulations
we have computed the data density matrix $R_{y}(b)$ and applied the
algorithm with different degrees of freedom:

\begin{enumerate}

   \item  First, we assume that the mixing matrix, 
     $A$, is known.  We construct the $A$ matrix
     from the exact electromagnetic spectrum of the anisotropies of
     each component and fix it in the algorithm. 
     Therefore, we take as parameters for the fit  $\theta(b)=\{R_s(b),{\rm
       diag}(R_n(b))\}$ for each bin $b$. With this test we want
     to check the spatial separability of the components. In the following, we refer to this type of
     component separation as {\it A-fixed} separation.

   \item Secondly, we have performed what we call a {\it CMB semi-blind} separation. 
     The $A$ matrix is fitted
     as well as $R_s$ and $R_n$. but assuming a prior on the CMB
     electromagnetic spectrum. Thus, the columns of the $A$ matrix
     corresponding to the CMB are fixed to unity and the set of
     parameters for the fit is
     $\theta(b)=\{A_{i,j\ne {\rm CMB}},R_s(b),{\rm
       diag}(R_n(b))\}$ for each bin $b$. The initialization of $A$, except for the
      CMB, is performed as described in section~\ref{initialization}.
     This kind of prior in the CMB electromagnetic spectrum is a reasonable
     approximation because the former is well-known.
   \item Finally we have performed a {\it blind} separation fitting all the parameters
    $\theta(b)=\{A,R_s(b),{\rm diag}(R_n(b))\}$ for each bin $b$ including the
    CMB electromagnetic spectrum.
\end{enumerate}

We have performed these three types of analysis in all the
\emph{simplified} simulated sets. 
To ensure the reliability of the results we performed 10000 EM iterations 
and checked, for each simulation, the convergence
of the EM algorithm.
In the following we present the main results obtained in reverse
order going from (iii) to (i) and if not stated otherwise we
assume white noise $R_{n}(b) = \sigma^{2}_{TT,EE,BB}$

%~~~~~~~~~~~~~~~~~~~~~~~~~~~~~~~~~~~~~~~~~~~~~~~~~~~~~~~~~~~~~~~~~~~~~~~~~~~~~~~~~~~~
\subsection[]{{\it Blind} separation analysis}
\label{blindsep}
We present here a {\it blind} analysis of the {\bf \tt [planck~c]} simulations
for which we assume three physical components in the data: CMB 
and \emph{simplified}-dust and synchrotron emissions.
The noise and physical-components density matrices 
as well as the mixing matrix are initialized as described in section~\ref{initialization}.
No physical priors are assumed neither for synchrotron nor dust. 
For CMB, if not stated otherwise, we initialize the electromagnetic 
spectrum to 1 for temperature and polarization. This is a reasonable approximation
as we expect the Planck data to be calibrated to better than 1 \% \cite{bluebook}.

\subsubsection*{Reconstruction of the power spectra}
\begin{figure*}
\centering
 \includegraphics[width=17cm,height=10cm]{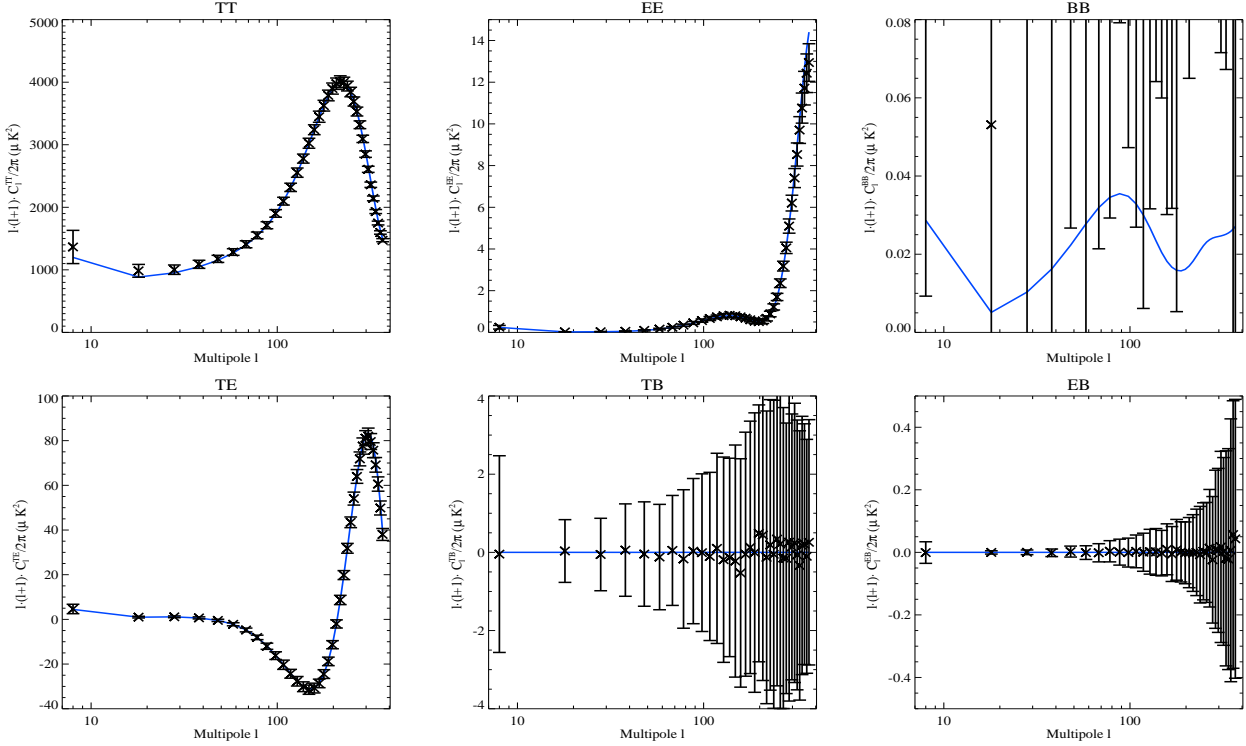}
%\plotancho{./CMB_spect_CMB_SYNC_DUST_r_0.7_nside_128_bin10_10000_colorless_cmb_ini.ps}
\caption[]{{\it Blind} reconstructed CMB power spectra (crosses),
$C_\ell^{TT}$, $C_\ell^{EE}$, $C_\ell^{BB}$, $C_\ell^{TE}$,
$C_\ell^{TB}$ and $C_\ell^{EB}$, in $\mu {\rm K}_{\rm CMB}^2$, at 100 GHz
for the {\bf \tt [planck~c]} simulations. We overplot
in blue the input model. Error bars are computed using
a total of 300 simulations.}
\label{fig:planckcmbfgd128}
\end{figure*}

Figure~\ref{fig:planckcmbfgd128} shows the {\it blind} reconstructed CMB temperature 
and polarization power spectra, in $\mu$K$^{2}_{\rm CMB}$, for the 100~GHz channel (crosses). 
The input model is overplotted in blue. 
The mean and the error bars are computed from the analysis of 300 simulations.

%In order to test the performance of the MD-MC separation method to
%polarized data we have simulated CMB and noise full sky maps for the
%Planck experiment ({\bf \tt [planck~b]} with $r=0.7$). Results for the
%CMB are displayed in fig. \ref{fig:planckcmb128}. Reconstructed
%spectra are compared to the input model and error bars are the
%dispersion over $\simeq$ 150 realizations we performed. Note that this
%separation is a semi-blind separation because here, the $A$ matrix
%contains only the CMB electromagnetic spectra.

We observe that $C^{TE}_\ell$, $C^{TB}_\ell$
and $C^{EB}_\ell$ are recovered with no bias, up to $\ell = 383$, which
is the largest accessible $\ell$ value at
$n_{\rm nside}=128$. In the same way, $C^{EE}_\ell$ is accurately reconstructed
except for the very high $\ell$ values for which pixelization
problems may appear (see section~\ref{sec:smallscales}). We also reconstruct efficiently  
$C^{TT}_\ell$ although a small bias (below 10 \%) is introduced
at low $\ell$ mainly due to confusion with the synchrotron emission
as discussed in the following. The $C^{BB}_\ell$  spectrum is 
not recovered at all and a significant bias is observed.
This bias, due mainly to statistical residual noise as discussed in the following
section, depends only on the signal to noise ratio and does not affect the 
reconstruction of the other components.

\begin{figure*}
\centering
%\plotancho{./DUST_spect_CMB_SYNC_DUST_r_0.7_nside_128_bin10_10000_colorless.ps}
 \includegraphics[width=17cm,height=10cm]{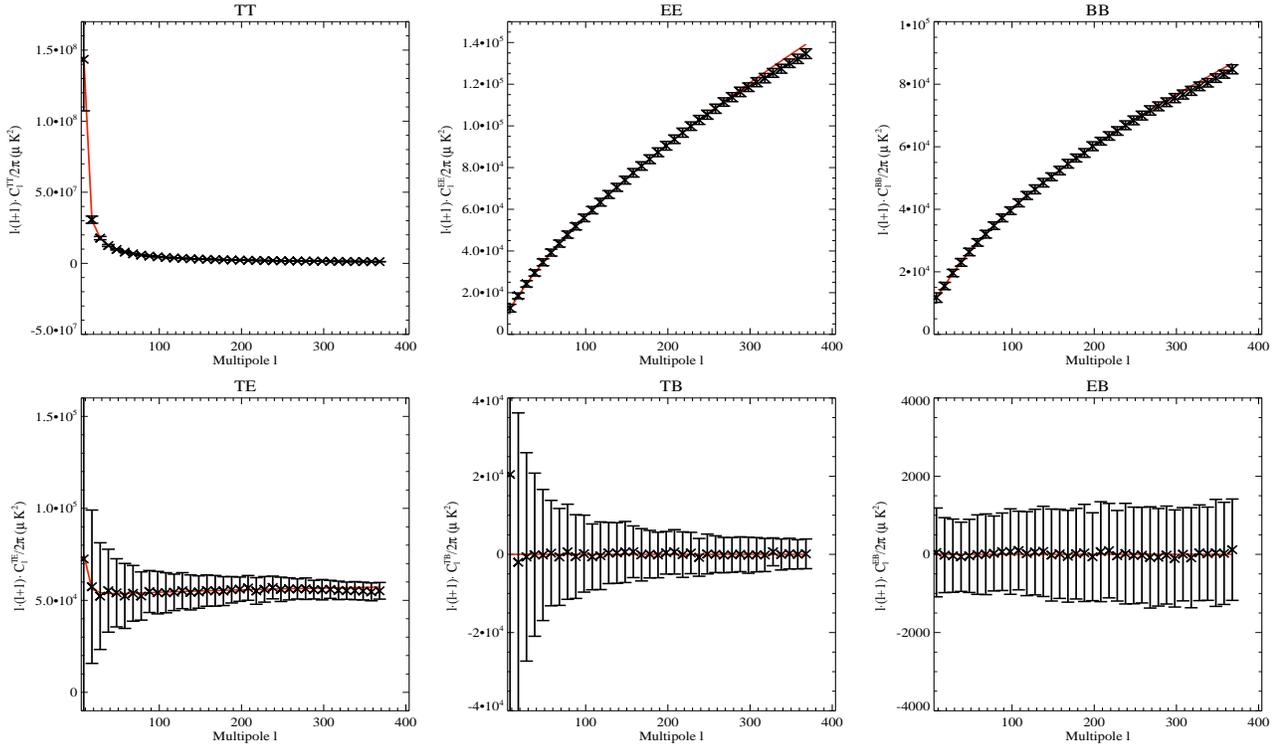}
\caption[]{{\it Blind} reconstructed \emph{simplified}-dust emission
power spectra, $C_\ell^{TT}$, $C_\ell^{EE}$, $C_\ell^{BB}$,
$C_\ell^{TE}$, $C_\ell^{TB}$ and $C_\ell^{EB}$, in $\mu
K_{\rm CMB}^2$ , at 353 GHz for the {\bf \tt [planck~c]}
simulations.  We overplot in red the input model. 
Error bars are computed using a total of 300 simulations.}
\label{fig:planckdustfgd128}
\end{figure*}

Figure~\ref{fig:planckdustfgd128} shows the reconstructed 
  \emph{simplified}-dust emission power spectra,  $C_\ell^{TT}$, $C_\ell^{EE}$, $C_\ell^{BB}$,
$C_\ell^{TE}$, $C_\ell^{TB}$ and $C_\ell^{EB}$, in $\mu
K_{\rm CMB}^2$, at 353 GHz.
For comparison we overplot in red the input model. The mean and
the error bars displayed were computed using a total of 
300 simulations. The reconstruction is fully efficient for
$C^{TT}_\ell$, $C^{TE}_\ell$,
$C^{TB}_\ell$ and $C^{EB}_\ell$ up to $\ell = 383$. 
The $C^{EE}_\ell$ and  $C^{BB}_\ell$ are accurately reconstructed
except at $\ell > 350$ where a small bias (below 10 \%) appears.
The $TB$ and $EB$ spectra are compatible with zero as expected
from the input model. These results are consistent with the fact
that the simulated \emph{simplified}-dust emission dominate the simulated maps
at the HFI channels for which the signal to noise ratio is larger.

\begin{figure*}
\centering
%\plotancho{./SYNC_spect_CMB_SYNC_DUST_r_0.7_nside_128_bin10_10000_colorless.ps}
 \includegraphics[width=17cm,height=10cm]{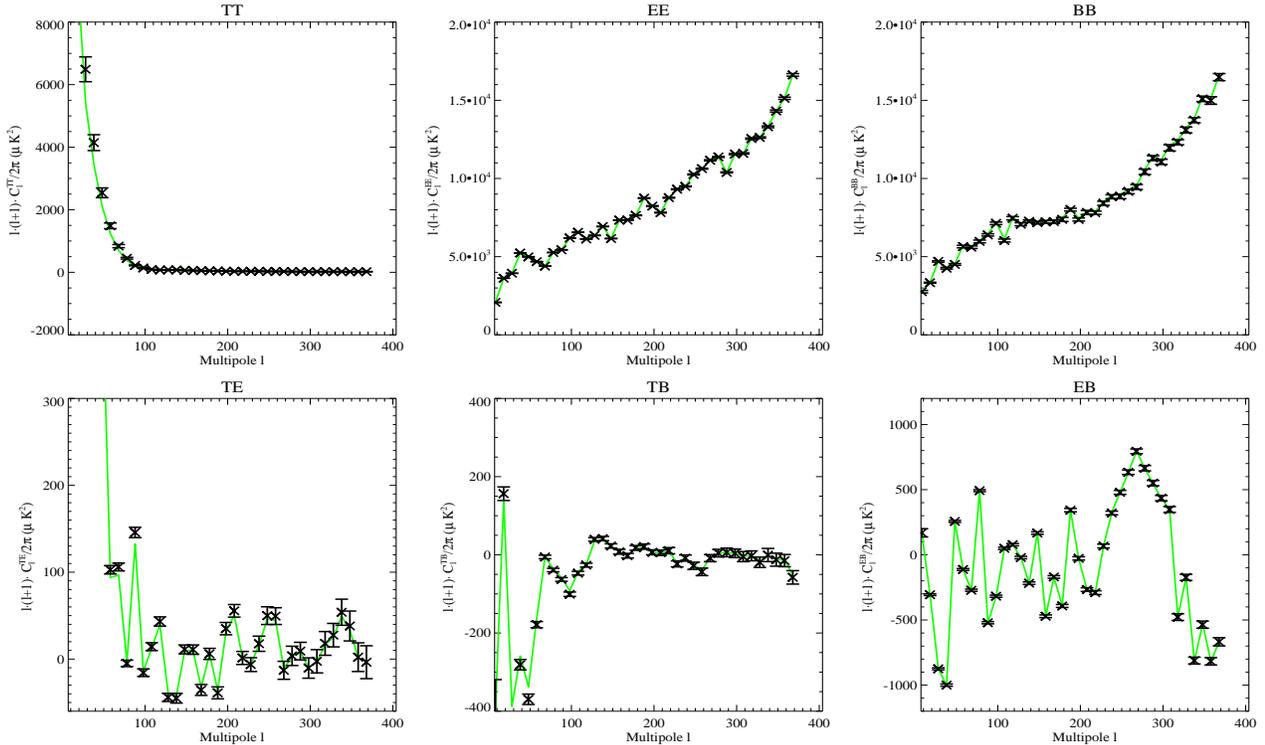}
\caption[]{Blind reconstructed synchrotron emission
power spectra, $C_\ell^{TT}$, $C_\ell^{EE}$, $C_\ell^{BB}$,
$C_\ell^{TE}$, $C_\ell^{TB}$ and $C_\ell^{EB}$, in $\mu
K_{\rm CMB}^2$ , at 353 GHz. We overplot in green the input model.
The error bars are computed using 300 simulations.
These are significantly smaller than for the other
two components because they do not include cosmic variance.
Indeed we use the same synchrotron temperature
and polarization maps for all the simulations. 
\label{fig:plancksyncfgd128}}
\end{figure*}

Finally, we present in figure~\ref{fig:plancksyncfgd128} the {\it blind} reconstructed
synchrotron power spectra, in $\mu
K_{\rm CMB}^2$, at 30~GHz.
We overplot in green the power spectrum of the input temperature
and polarization synchrotron map from \cite{giardino}.
Here again $C^{EE}_\ell$, $C^{BB}_\ell$, $C^{TE}_\ell$, $C^{TB}_\ell$ and $C^{EB}_\ell$ 
are recovered efficiently. A bias at low $\ell$ (below 20 \%) is observed for $C^{TT}_\ell$.
This is due, as discussed in the following, to the slight mixing-up of the synchrotron and CMB
emissions in temperature.

\subsubsection*{Reconstruction of the mixing matrix}

\begin{figure*}
\centering
        \includegraphics[width=17cm,height=4cm]{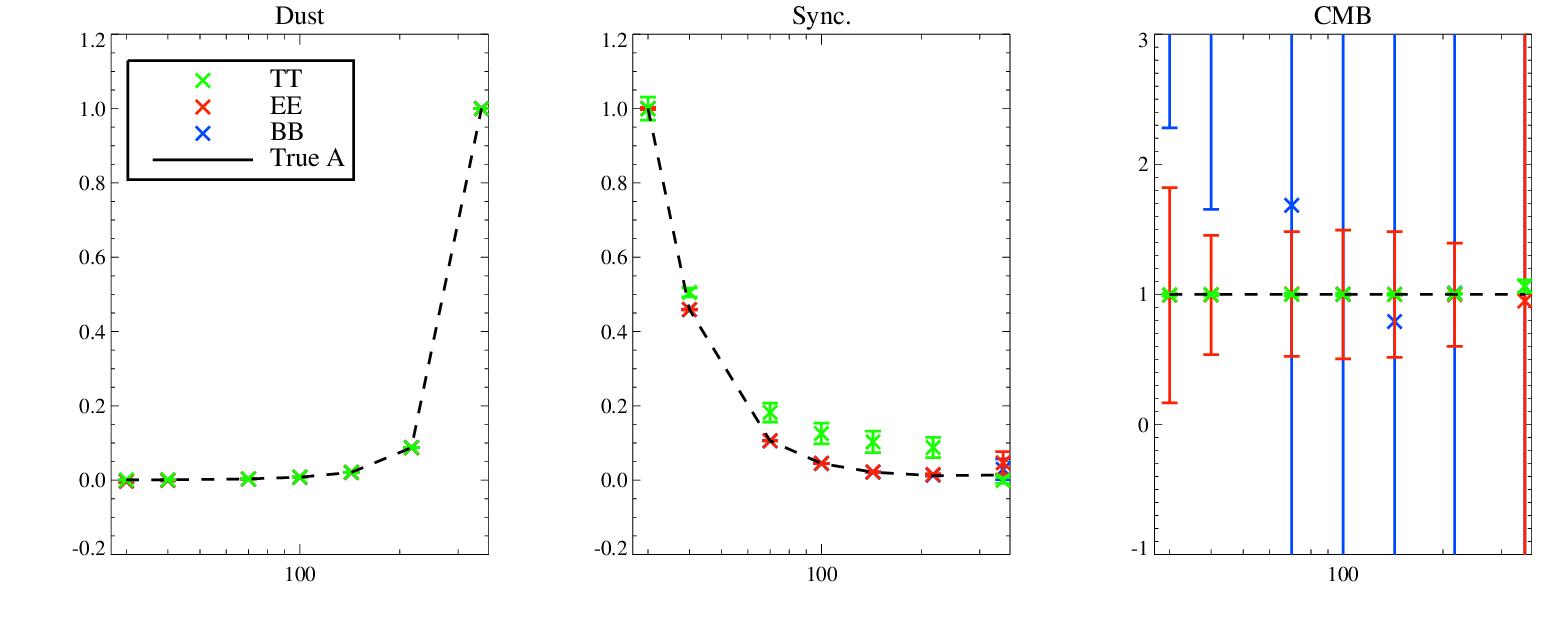}
        \includegraphics[width=17cm,height=4cm]{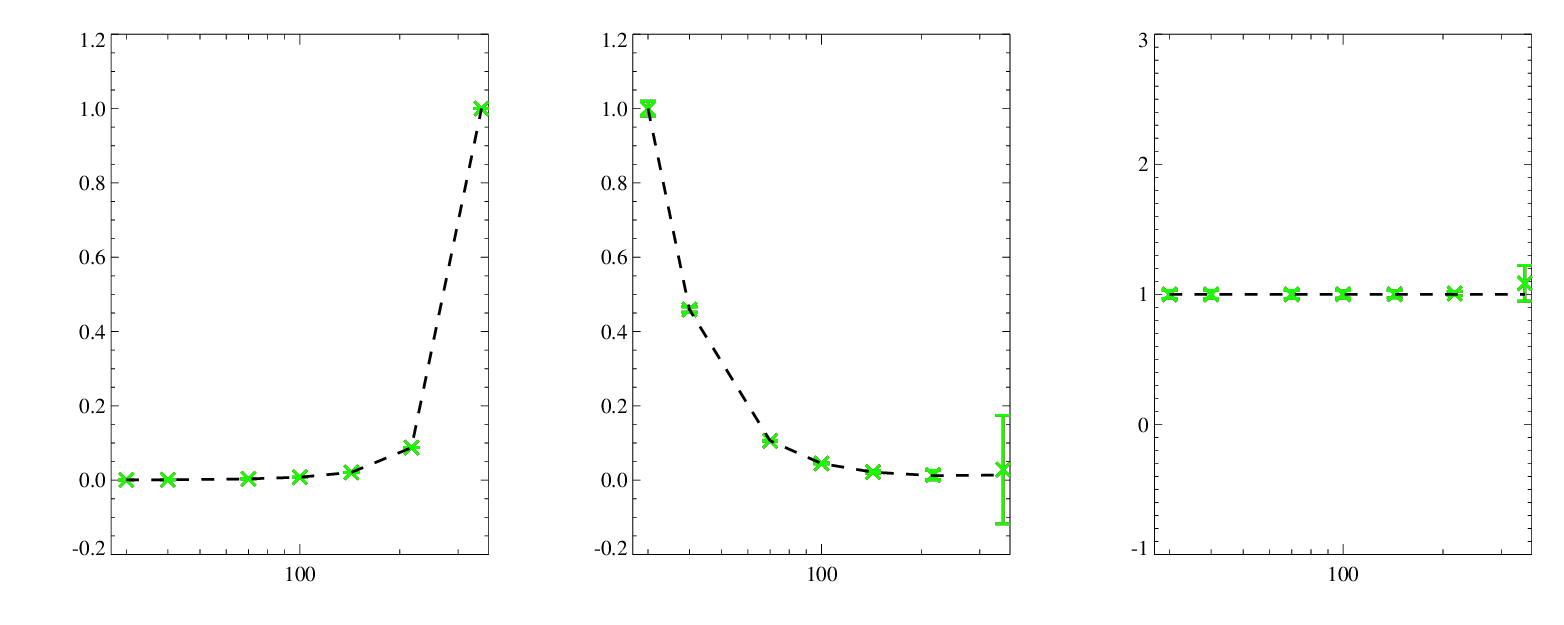}
        \includegraphics[width=17cm,height=4cm]{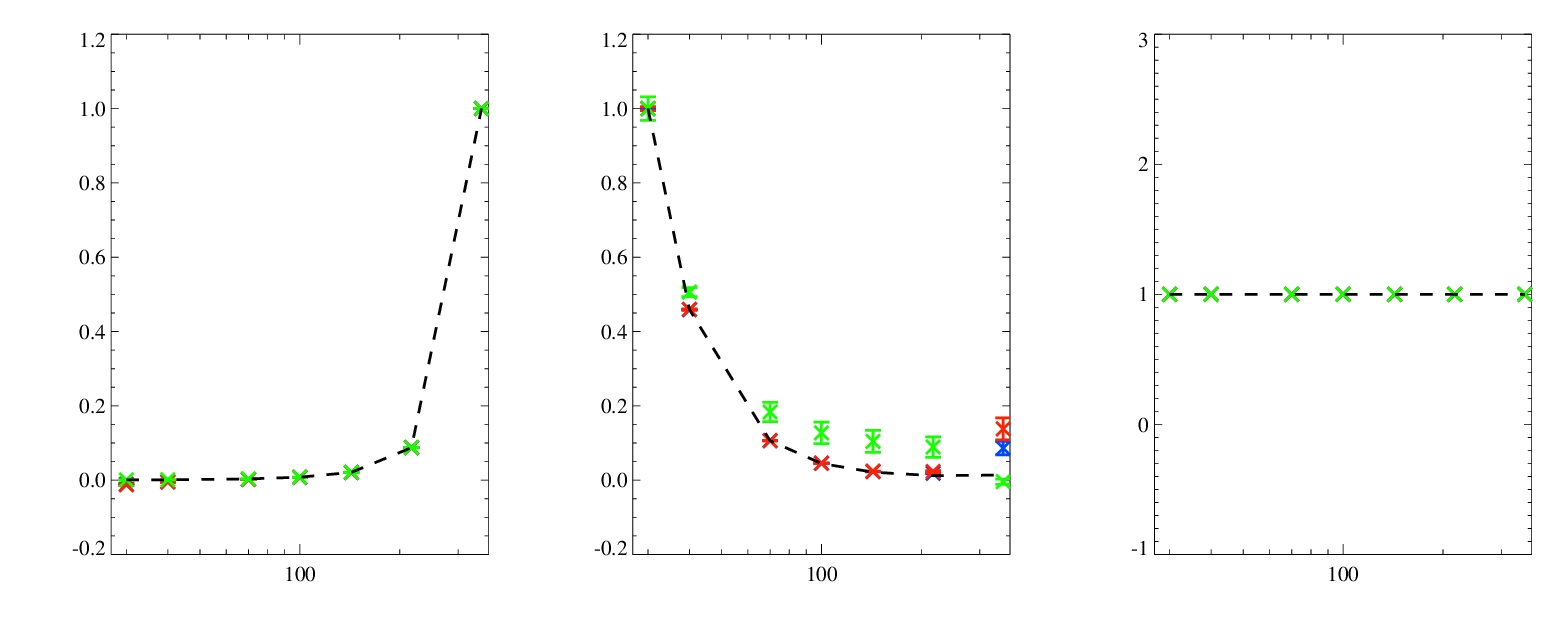}

\caption[fig:figure1]{Reconstructed mixing matrix, $A$.
  Columns, from left to right, correspond to the
  reconstructed $A$ matrix for dust, synchrotron and CMB respectively.
  Rows, from top to bottom, correspond to the {\it blind}, {\it blind} assuming
  equal temperature and polarization electromagnetic spectrum and
  the {\it CMB semi-blind} analysis of the {\bf \tt [planck~c]} simulations.
  We display the output electromagnetic spectrum computed for the $T$ (green),
  $E$ (red) and $B$ (blue) modes as well as the input electromagnetic
  spectrum (dashed black).
\label{fig:amat128}}
\end{figure*}

The slight mixing up between synchrotron and
CMB is better observed in the reconstructed mixing matrix.
The first row of figure \ref{fig:amat128} shows 
the {\it blind} recovered $A$ matrix for dust, synchrotron
and CMB in the case of the {\bf \tt [planck~c]} simulations. The electromagnetic spectrum for $T$, $E$
and $B$ are respectively traced in green, red and blue. For comparison we overplot
the input electromagnetic spectrum for each of the components (black dashed
line). For convenience we have renormalized the electromagnetic
spectrum so that it is unity at 353~GHz, 30~GHz and 100~GHz for dust, synchrotron
and CMB respectively. It is important to remark that the reconstruction
of the electromagnetic spectra for temperature and polarization
is performed independently. We observe that
the error bars are larger for polarization than for temperature
as we would expect from the smaller signal to noise ratio in polarization.

The \emph{simplified}-dust electromagnetic spectrum is reconstructed with no bias
both in temperature and polarization even at the lowest LFI
frequency channels. Furthermore, the reconstructed 
synchrotron electromagnetic spectra in polarization
are not biased. In temperature we observe that the
spectrum flattens out at intermediate frequencies between
70 and 217~GHz. This is the cause of the slight mixing up between
synchrotron and CMB.
This mixing up does not happen in polarization for which
the synchrotron emission dominates over the CMB emission.
Finally, the reconstruction of the CMB electromagnetic spectrum
from the $T$ and $E$ modes, although noisy for the latter, is not biased. However for the
$B$ modes the reconstruction is very poor because of the very low 
signal to noise ratio (below $10^{-2}$ for $\ell > 100$).

We have repeated the analysis with no prior in the electromagnetic spectrum
of the CMB. We use instead the eigenvector corresponding to
the third larger eigenvalue of the data density matrix. 
The results for dust and synchrotron remain unchanged. For CMB
the electromagnetic spectrum at 30 and 353~GHz is not
reconstructed neither in temperature nor in polarization and
the results for $B$ are significantly degraded at all frequencies. 
However, the results on the reconstruction of the
spatial power spectra remain unchanged for all the
physical components including CMB. This can be easily
understood as the reconstruction of the CMB power spectra is
mainly dominated by the intermediate frequency maps, from 70 to 217~GHz,
where the $A$ matrix is accurately reconstructed.

\subsubsection*{Assuming equal temperature and polarization electromagnetic spectrum}

In the previous analysis we have computed the electromagnetic spectrum of the
physical components independently for each mode $T$, $E$ and $B$. In a more realistic
approach we should consider a single electromagnetic spectrum for the polarization
$E$ and $B$ modes which may be different from the temperature one. 
We have repeated the analysis under the above  hypothesis and 
the results remain roughly the same with respect to the reconstruction of
the spatial power spectra and of the electromagnetic spectrum.
For synchrotron and \emph{simplified}-dust the $E$ and $B$ modes have roughly 
the same power in our simulations and
therefore we expect only variations in the error bars.
For CMB the $E$ mode largely dominates the $B$ mode and therefore we expect
no significant contribution from the latter to the electromagnetic spectrum
reconstruction.

\begin{figure*}
\centering
 \includegraphics[width=17cm,height=10cm]{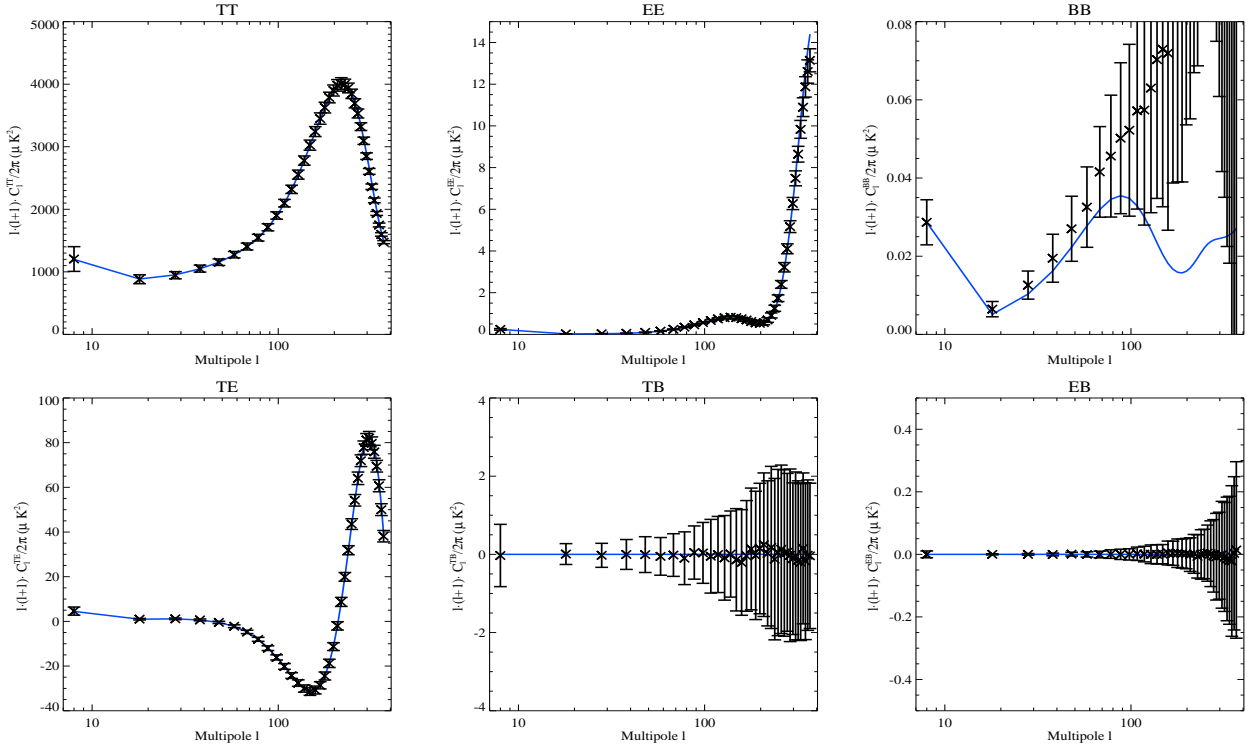}
\caption[]{Same as figure~\ref{fig:planckcmbfgd128} except for a {\it blind} analysis assuming equal temperature
and polarization electromagnetic spectrum.}
\label{fig:tpqspectplanckcmbfgd128}
\end{figure*}

\begin{figure*}
\centering
 \includegraphics[width=17cm,height=10cm]{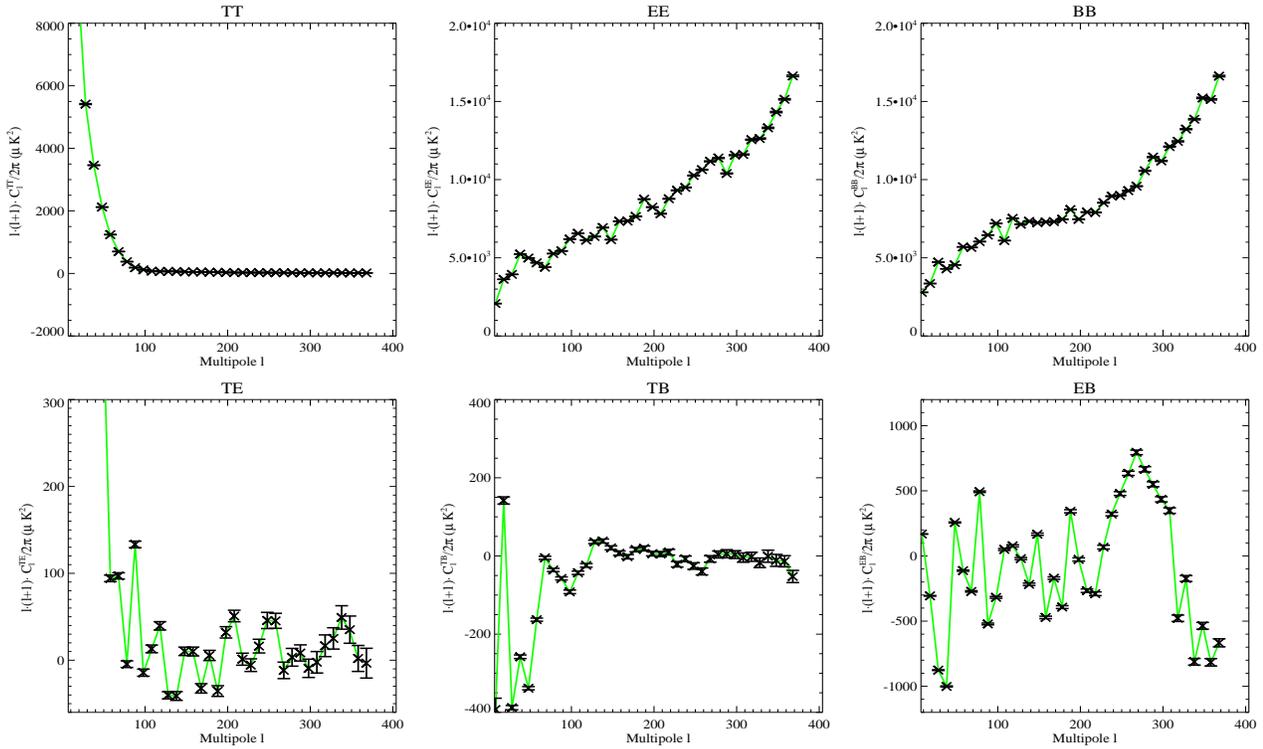}
\caption[]{Same as figure~\ref{fig:plancksyncfgd128} but for a {\it blind} analysis 
assuming equal temperature and polarization electromagnetic spectrum.
\label{fig:tpqspectplancksyncfgd128}}
\end{figure*}

The differences between the temperature and polarization electromagnetic
spectra are expected to be small for dust and synchrotron \cite{bluebook} and none for
the CMB \cite{zaldarriaga}. Therefore, in the case of a perfectly calibrated
experiment, we can consider, in a first approximation, that the polarization and the
temperature electromagnetic spectra are the same. The {\it blind} analysis of the
{\bf \tt [planck~c]} simulations under this hypothesis shows no evidence of mixing up
between synchrotron and CMB. This can be clearly observed in the second row of
figure~\ref{fig:amat128} where we represent from left to right the
reconstructed electromagnetic spectrum of the dust, synchrotron and CMB 
emissions respectively. No bias is observed for any of the physical components
including synchrotron for which the flatten out of the spectrum observed before
is not present.

Figures~\ref{fig:tpqspectplanckcmbfgd128} and \ref{fig:tpqspectplancksyncfgd128} show 
the CMB and synchrotron reconstructed spatial power spectra in temperature and polarization. 
We do not observe a bias neither on the synchrotron nor on the CMB $TT$ power spectra. 
Furthermore, we observe that
the CMB  $BB$ modes, although biased at large $\ell$ values, are fairly reconstructed
up to $\ell = 60$.
For the other modes, the results are similar to those
presented in the previous section. 
%From these results we can conclude that the bias in the
%synchrotron and CMB $TT$ modes was due to the bad estimation of the synchrotron 
%electromagnetic spectrum. In the same way, the poor reconstruction of the CMB $BB$ modes
%was due to the bad estimation of the CMB $B$ electromagnetic spectrum.

\subsection{{\it Semi-blind} separation}
\begin{figure*}
\centering
 \includegraphics[width=17cm,height=10cm]{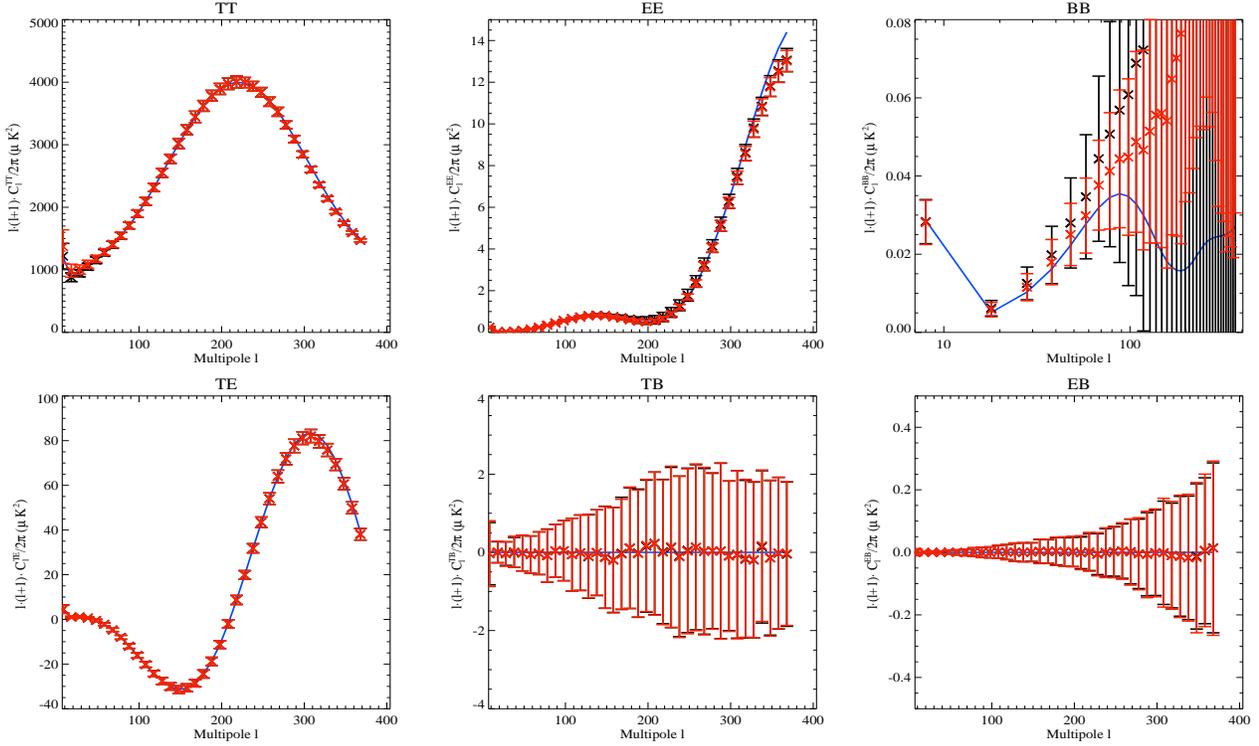}
\caption[]{Same as figure~\ref{fig:planckcmbfgd128} but for a CMB {\it semi-blind} (black) and
{\it A-fixed} (red) analysis (see text for details).
\label{fig:semiblindplanckcmbfgd128}}
\end{figure*}

The separation method allows us to easily include previous knowledge on the physical components
either as priors or as facts. In the previous section we considered a prior on the CMB
emission. In the following we move a step forward in the analysis 
assuming the CMB electromagnetic spectrum
known and performing what we call a {\it CMB semi-blind} analysis. 
For this analysis, the columns of the mixing matrix corresponding to the CMB 
are initialized to the CMB electromagnetic spectrum and are not updated by the algorithm. 
For the other components we consider independent electromagnetic spectra for temperature
and polarization and they are initialized as for the {\it blind} analysis.

The third row of figure~\ref{fig:amat128} shows 
the reconstructed mixing matrix for the {\it CMB semi-blind} analysis considering the
{\bf \tt [planck~c]} simulations. The results are similar to those of the {\it blind} analysis
discussed before. The dust electromagnetic spectrum is accurately recovered in temperature
and polarization. For synchrotron, the polarization spectrum is accurately recovered
but the temperature one flattens out at intermediate frequencies with respect to
the input model. Therefore, the reconstructed
\emph{simplified}-dust and synchrotron spatial power spectra in temperature and polarization are similar to those
of the {\it blind} analysis. The \emph{simplified}-dust power spectra are accurately reconstructed in temperature
and polarization. For synchrotron the power spectra are also accurately reconstructed except for the
$TT$ mode which present a slight bias (below 20\%) at large angular scales ($\ell < 60$).
In general the error bars are smaller for the {\it CMB semi-blind}
analysis.
% as discussed
%in section~\ref{sec:foregroundeffects}. 

The reconstructed CMB power spectra for the {\it CMB semi-blind} analysis 
are represented in black on figure~\ref{fig:semiblindplanckcmbfgd128}. 
All of them are accurately reconstructed with no bias
except for the $BB$ mode. For the latter the reconstruction is accurate up to $\ell = 60$ and
there on is biased. This bias is due to residual noise and is not related to the uncertainties
on the reconstruction of the electromagnetic spectrum for the other physical components.
To check this we have also performed a {\it A-fixed} analysis assuming the electromagnetic
spectrum of all physical components known. The results of this analysis are overplotted in 
red on the figure. We observe that reconstruction is equivalent to that of the
CMB {\it semi-blind} analysis but for the error bars which are smaller
%as described in section~\ref{sec:foregroundeffects}. 
This indicates 
that the bias in the $BB$ mode is mainly due to residual noise as
discussed in section~\ref{sec:colornoisemodel}.

% Start of old part discussions

\label{sec:discusions}
%~~~~~~~~~~~~~~~~~~~~~~~~~~~~~~~~~~~~~~~~~~~~~~~~~~~~~~~~~~~~~~~~~~~~~~~~~
\subsection[]{Reconstruction of the small angular scales}
\label{sec:smallscales}
\begin{figure*}
\centering
 \includegraphics[width=17cm,height=10cm]{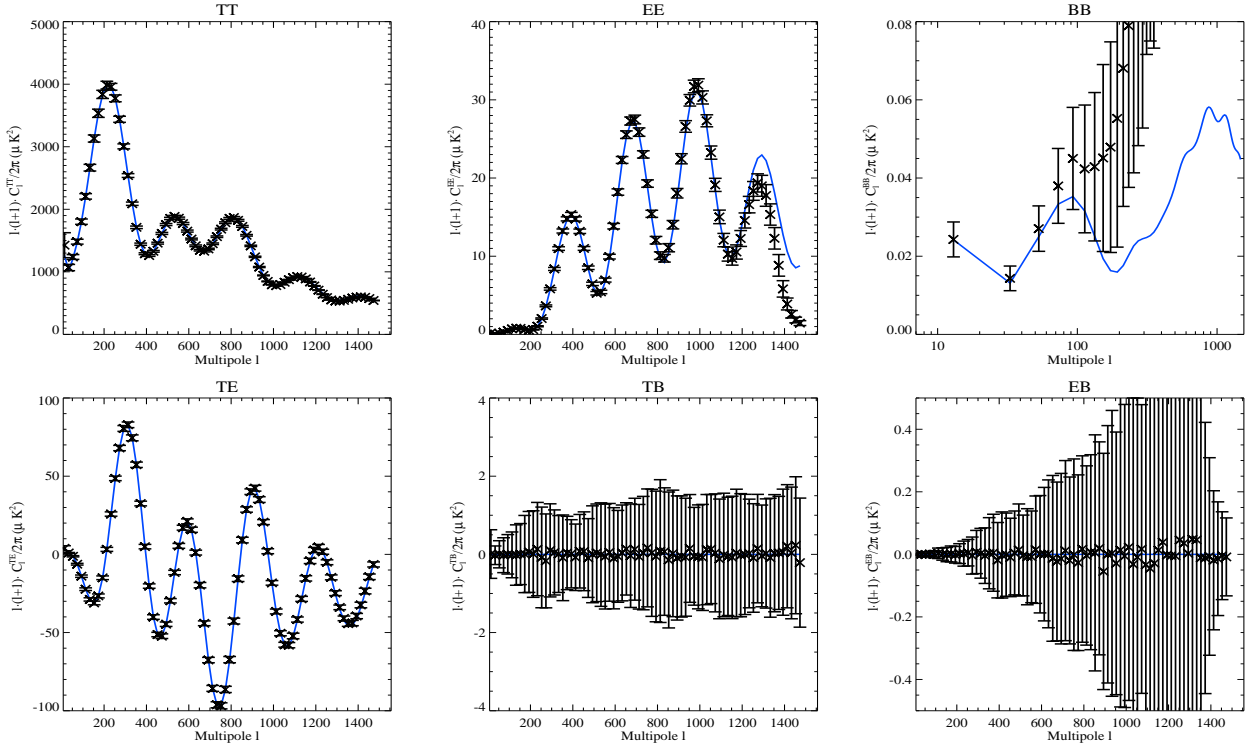}
\caption[fig:figure1]{Same as figure~\ref{fig:planckcmbfgd128} but for
  the {\bf \tt [planck~a]} simulations for a {\it CMB
    semi-blind} analysis.
\label{fig:cmbfgd512planck}}
\end{figure*}

\begin{figure*}
\centering
 \includegraphics[width=17cm,height=10cm]{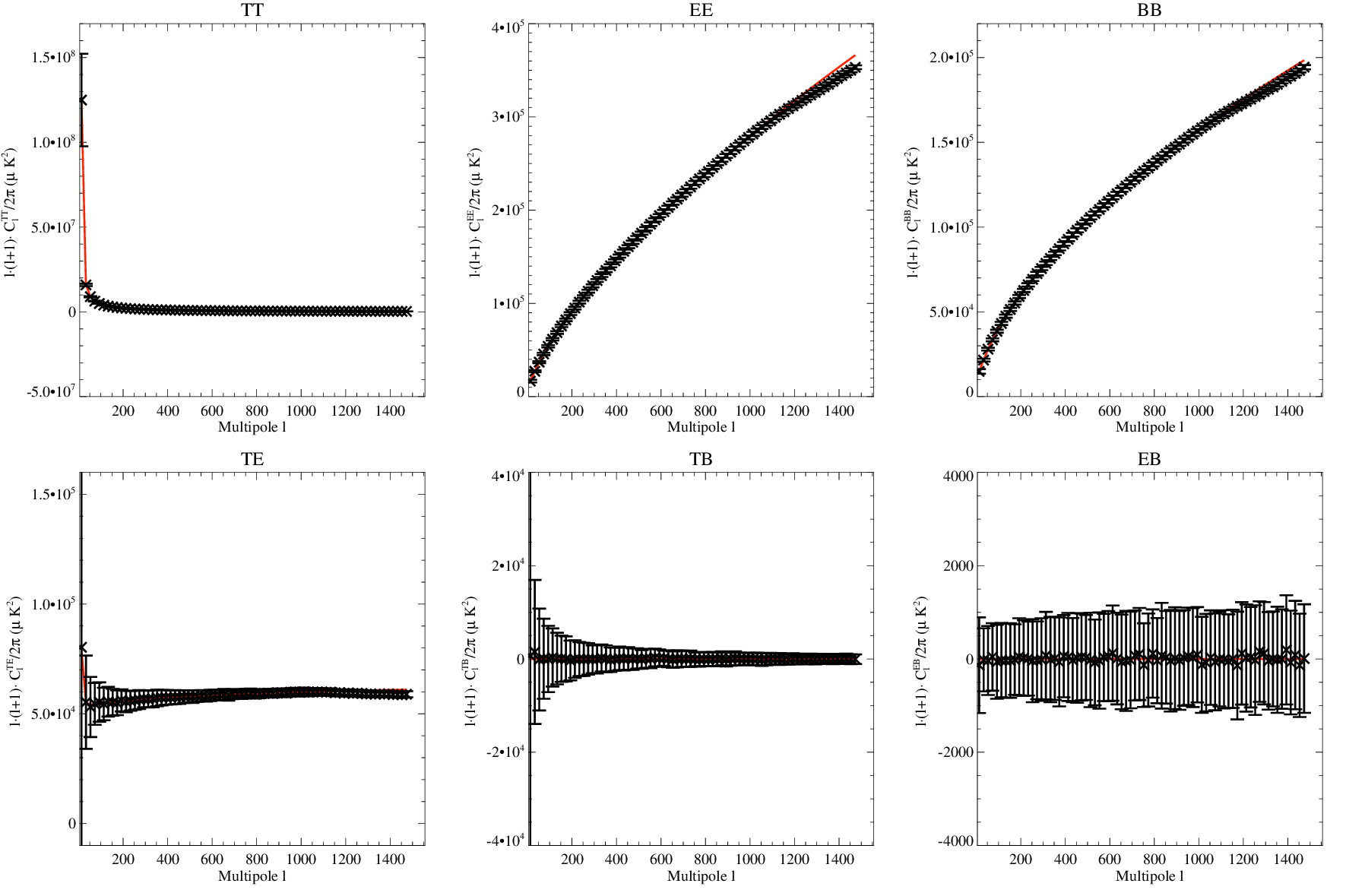}
\caption[fig:figure1]{Same as figure~\ref{fig:planckdustfgd128} but for the {\bf \tt [planck~a]} simulations for a {\it CMB
    semi-blind} analysis.
\label{fig:dustfgd512planck}}
\end{figure*}

\begin{figure*}
\centering
 \includegraphics[width=17cm,height=10cm]{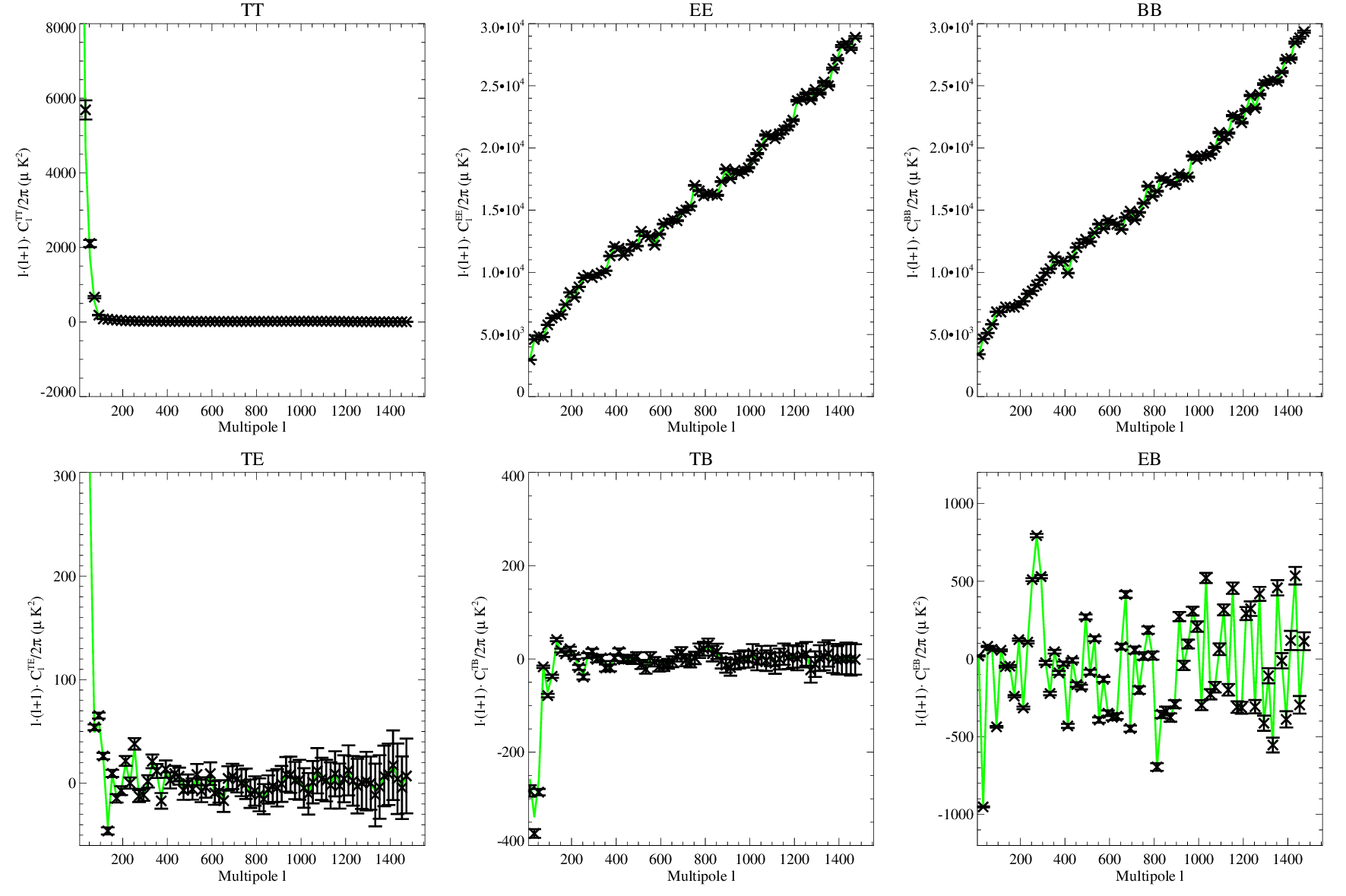}
\caption[fig:figure1]{Same as figure~\ref{fig:plancksyncfgd128} but for the {\bf \tt [planck~a]} simulations 
for a {\it CMB semi-blind} analysis. \label{fig:syncfgd512planck}}
\end{figure*}

In the previous section we have fully described the analysis of the
{\bf \tt [planck~c]} simulations at $n_{\rm side}=128$ for which the
reconstruction of the spatial power spectra was limited to
$\ell = 383$. In some cases we have observed small biases in the
polarization auto power spectra at large $\ell$ values which may be due
to pixelization problems (we exclude in here the bias observed
in the CMB $BB$ modes which is due to residual noise). To check this
hypothesis we have also performed the {\it blind}, {\it CMB semi-blind}
and {\it A-fixed} analysis on the {\bf \tt [planck~a]} simulations
for which we can reconstruct the angular power spectra up to $\ell = 1535$.
As the resolution of the Planck best channels is $\sim 5$ arcmin 
a more realistic analysis will require simulations at $n_{side} = 2048$
which are far too much time demanding for our computational capabilities.

The results obtained for the {\bf \tt [planck~a]} simulations are very similar
to those for the {\bf \tt [planck~c]} ones. For illustration 
figures~\ref{fig:cmbfgd512planck},~\ref{fig:dustfgd512planck} and~\ref{fig:syncfgd512planck}
show the reconstructed temperature and polarization power spectra for
CMB, \emph{simplified}-dust and
synchrotron in the case of a {\it CMB semi-blind} analysis. The
synchrotron and \emph{simplified}-dust power spectra
are accurately reconstructed in temperature and polarization. We observe a slight bias
in the dust $EE$ and $BB$ modes as for the {\bf \tt [planck~c]} simulations but at
much larger $\ell$ values. This indicates that this
bias is related to pixelization effects. The same effect is observed
for the CMB power spectra. The bias in the $EE$ modes is present at much larger $\ell$
than for the {\bf \tt [planck~c]} simulations. 
The reconstruction of the CMB $BB$ modes is accurate at low $\ell$ ($\ell < 60$) and present
a residual noise bias at large $\ell$ values as discussed in the previous
section.

%~~~~~~~~~~~~~~~~~~~~~~~~~~~~~~~~~~~~~~~~~~~~~~~~~~~~~~~~~~~~~~~~~~~~~~~~~
\subsection[]{Color noise model}
\label{sec:colornoisemodel}
\begin{figure*}
\centering
%\plotancho{./Noise_CMB_SYNC_DUST_r_0.7_nside_512_bin10_10000_ALLBLIND_test300_keep_at_min.ps}
 \includegraphics[width=17cm,height=5cm]{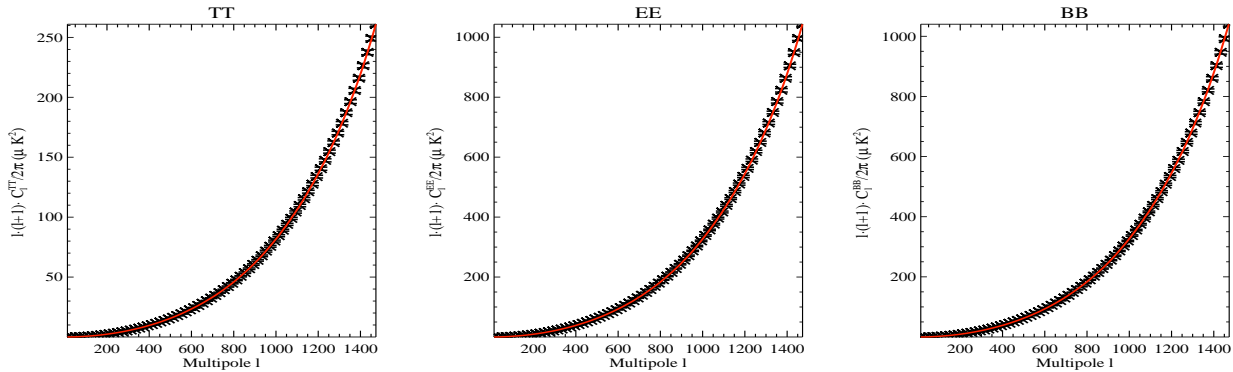}
\caption[fig:figure1]{Reconstructed noise angular auto power spectra (black),
   in $\mu {\rm K}_{\rm CMB}^2$, at 100 GHz for $T$, $E$ and
  $B$  for the {\bf \tt [planck~a]} simulations in the case of a {\it CMB semi-blind}
  separation. We overplot in red the noise auto power spectra at 100~GHz
  as obtained from 100 realizations of noise-only maps.
\label{fig:noise128}}
\end{figure*}

As seen before, the bias observed in the CMB $BB$ power spectrum is most
probably due to residual noise from the separation. Therefore, it is interesting to check both the accuracy
of the noise reconstruction and the limitations of the white noise model imposed.
With respect to the latter we have repeated all the analysis presented above assuming
a color noise model such that the noise power spectra are estimated for each bin $b$
in $\ell$. For the {\it blind} analysis the results are slightly worse in the sense that the mixing
up between CMB and synchrotron is more significant. This is not surprising since the noise
in the data is white and we are artificially reducing the number of degrees of freedom in
the fit. Actually, the mismatch between the data and the physical component power spectra 
can be compensated by changing the noise power spectra. In the case of the {\it CMB semi-blind}
and {\it A-fixed} analysis the results for the white and color noise model present
no significant differences.
    
Figure~\ref{fig:noise128} shows, in black, 
the reconstructed noise angular power
spectra, $C_\ell^{N,TT}$, $C_\ell^{N,EE}$, $C_\ell^{N,BB}$, in $\mu {\rm K}_{\rm CMB}^2$, 
at 100~GHz for the {\bf \tt [planck~a]} simulations in the case of a {\it CMB semi-blind}
analysis. We overplot, in black, the power spectra of the noise at 100~GHz
obtained from 100 realizations of noise-only maps. The noise spectra are reconstructed
down to the accuracy of the estimation of the input model both for $TT$ and $EE$, 
well below 10$^{-3}$~\%. For the $BB$ noise spectrum there is a small bias 
which is of the order of $5 \times 10^{-3}$~\% at $\ell =200$ and 
around $2 \times 10^{-2}$~\% at $\ell =1500$. 
 
Therefore, to improve the reconstruction of the CMB $BB$ modes we need a better estimation
of the $BB$ noise power spectrum. For this purpose we need to improve the likelihood maximization
algorithm. For temperature-only separation, \cite{delabrouille} complemented the EM algorithm with
a direct maximization of the likelihood function via a Newton-Raphson algorithm.
For polarization similar algorithms can be used but due to the degree of complexity of the
problem (6 correlated modes per physical component instead of 1 in the temperature only case) 
and for the sake of clarity these will be discussed in a forthcoming paper.

\section[]{Towards a more realistic sky model}

\label{realistic_model}

After testing intensively our algorithm on our \emph{simplified}
  model, deducting the global performances of the spectral matching
  reconstruction in temperature and polarization we are interested in
  the performances of the algorithm for significantly spatially
  correlated components. For this purpose, we have used the {\bf \tt
    [planck~d]} simulations.

\begin{figure*}
\centering
 \includegraphics[width=17cm,height=10cm]{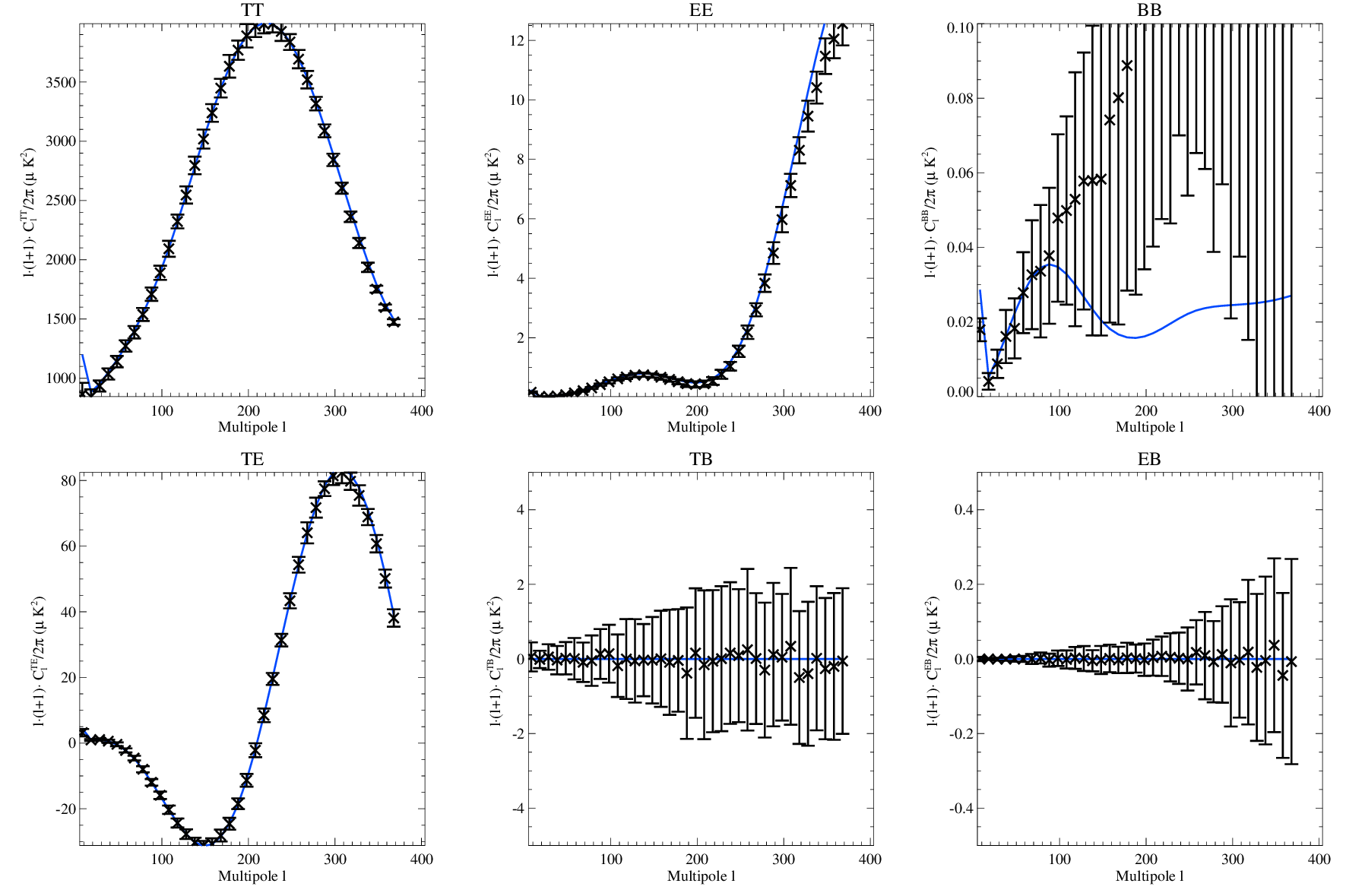}
\caption[fig:figure1]{\emph{A-fixed} reconstruction o the CMB $C_\ell^{TT}$, $C_\ell^{EE}$, $C_\ell^{BB}$, $C_\ell^{TE}$,
$C_\ell^{TB}$ and $C_\ell^{EB}$ power spectra (crosses),
 in $\mu {\rm K}_{\rm CMB}^2$, at 100
GHz in the case of the \emph{realistic} model with 4 components. We overplot
in blue the input model. Error bars are computed using
a total of 100 simulations. \label{4comp_realistic_cmb}}
\end{figure*}

\begin{figure*}
\centering
 \includegraphics[width=17cm,height=10cm]{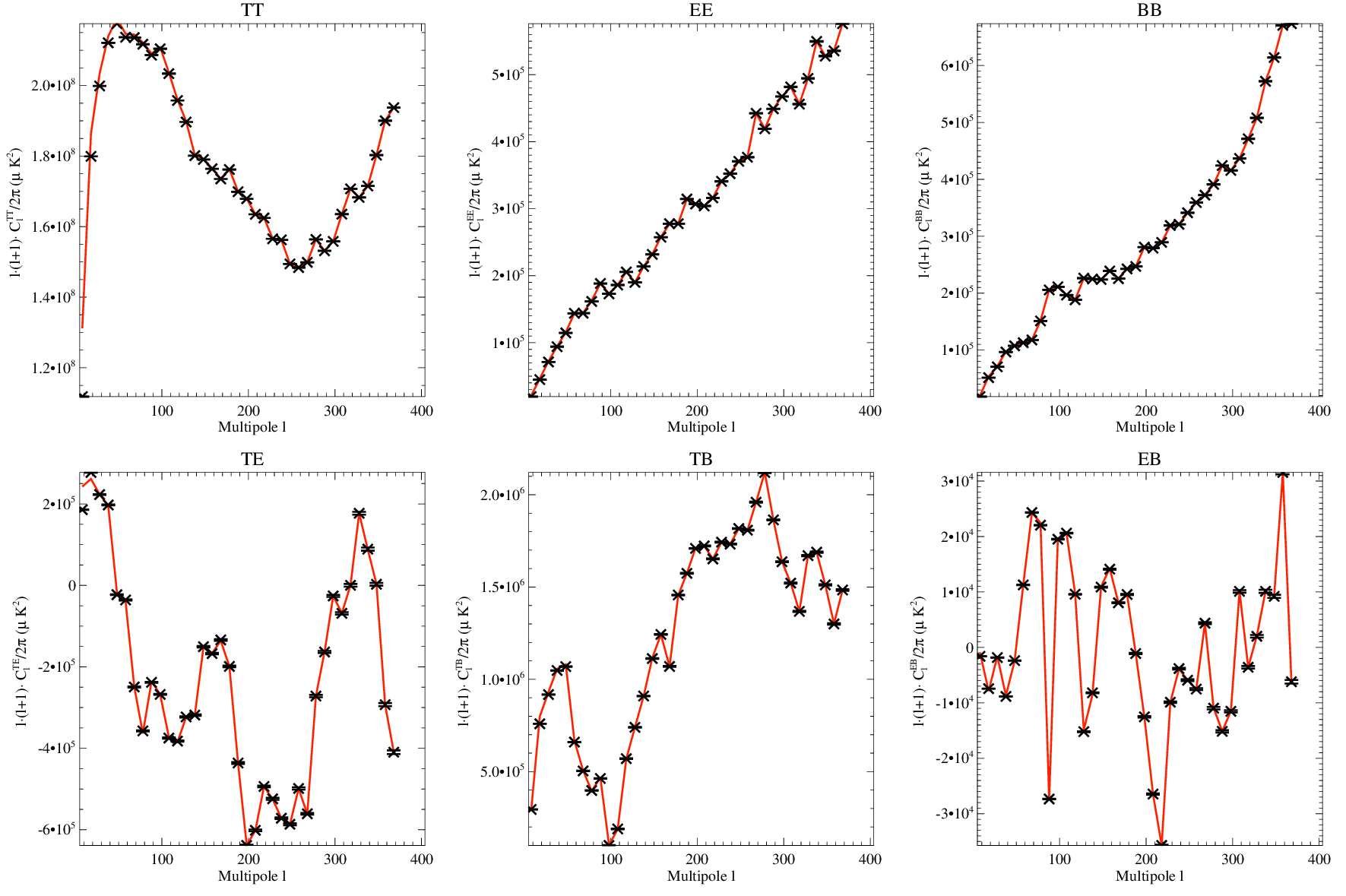}
\caption[fig:figure1]{\emph{A-fixed} reconstruction of the
  \emph{realistic} dust $C_\ell^{TT}$, $C_\ell^{EE}$, $C_\ell^{BB}$, $C_\ell^{TE}$,
$C_\ell^{TB}$ and $C_\ell^{EB}$ power spectra (crosses),
 in $\mu {\rm K}_{\rm CMB}^2$, at 353
GHz in the case of the \emph{realistic} model with 4 components. We overplot
in red the input model. Error bars are computed using
a total of 100 simulations. \label{4comp_realistic_dust}}
\end{figure*}

\begin{figure*}
\centering
 \includegraphics[width=17cm,height=10cm]{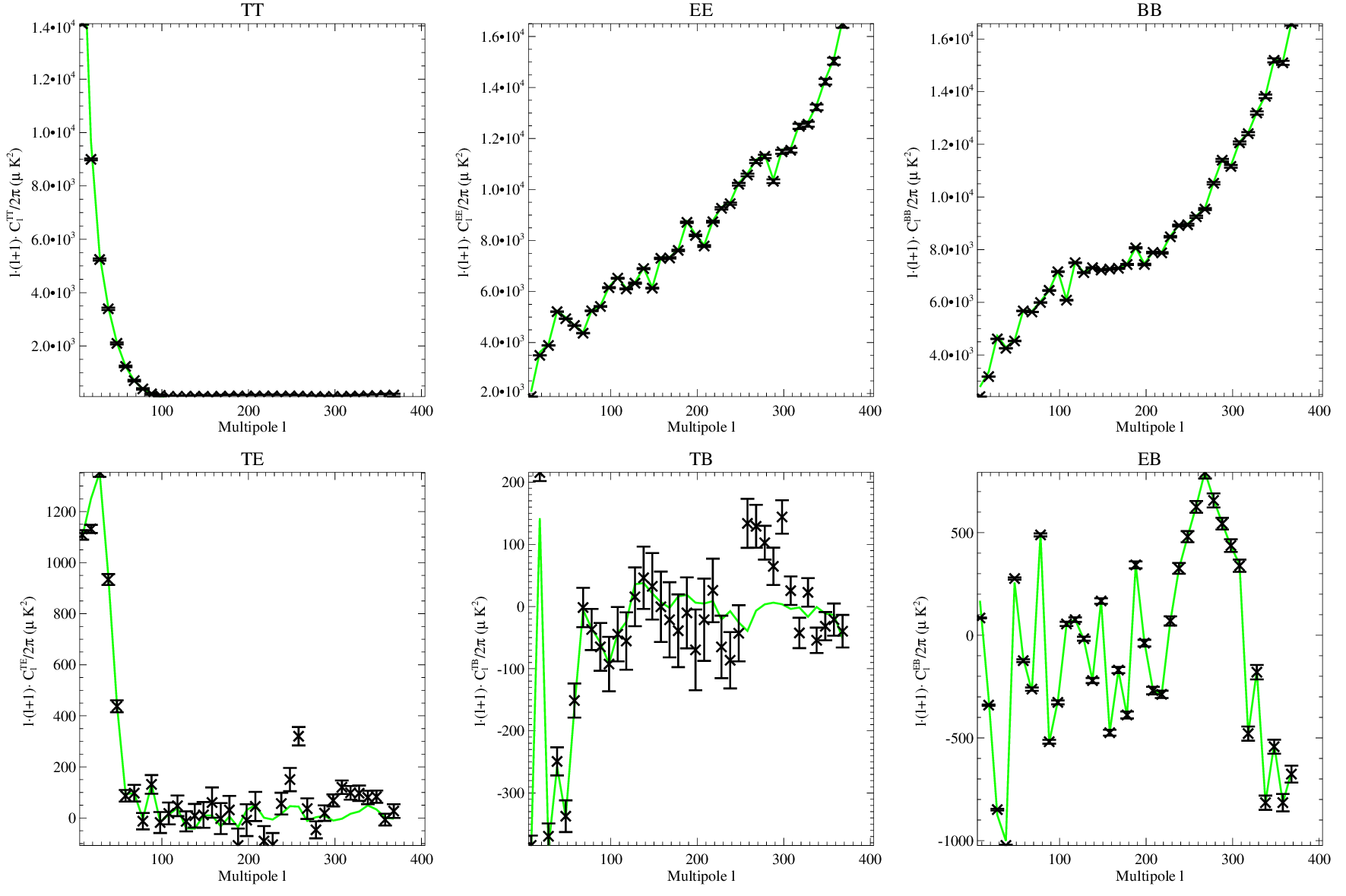}
\caption[fig:figure1]{\emph{A-fixed} reconstruction of the
  synchrotron emission $C_\ell^{TT}$, $C_\ell^{EE}$, $C_\ell^{BB}$, $C_\ell^{TE}$,
$C_\ell^{TB}$ and $C_\ell^{EB}$ power spectra (crosses),
 in $\mu {\rm K}_{\rm CMB}^2$, at 30
GHz in the case of the \emph{realistic} model with 4 components. We overplot
in green the input model. Error bars are computed using
a total of 100 simulations. \label{4comp_realistic_sync}}
\end{figure*}

\begin{figure*}
\centering
 \includegraphics[width=17cm,height=10cm]{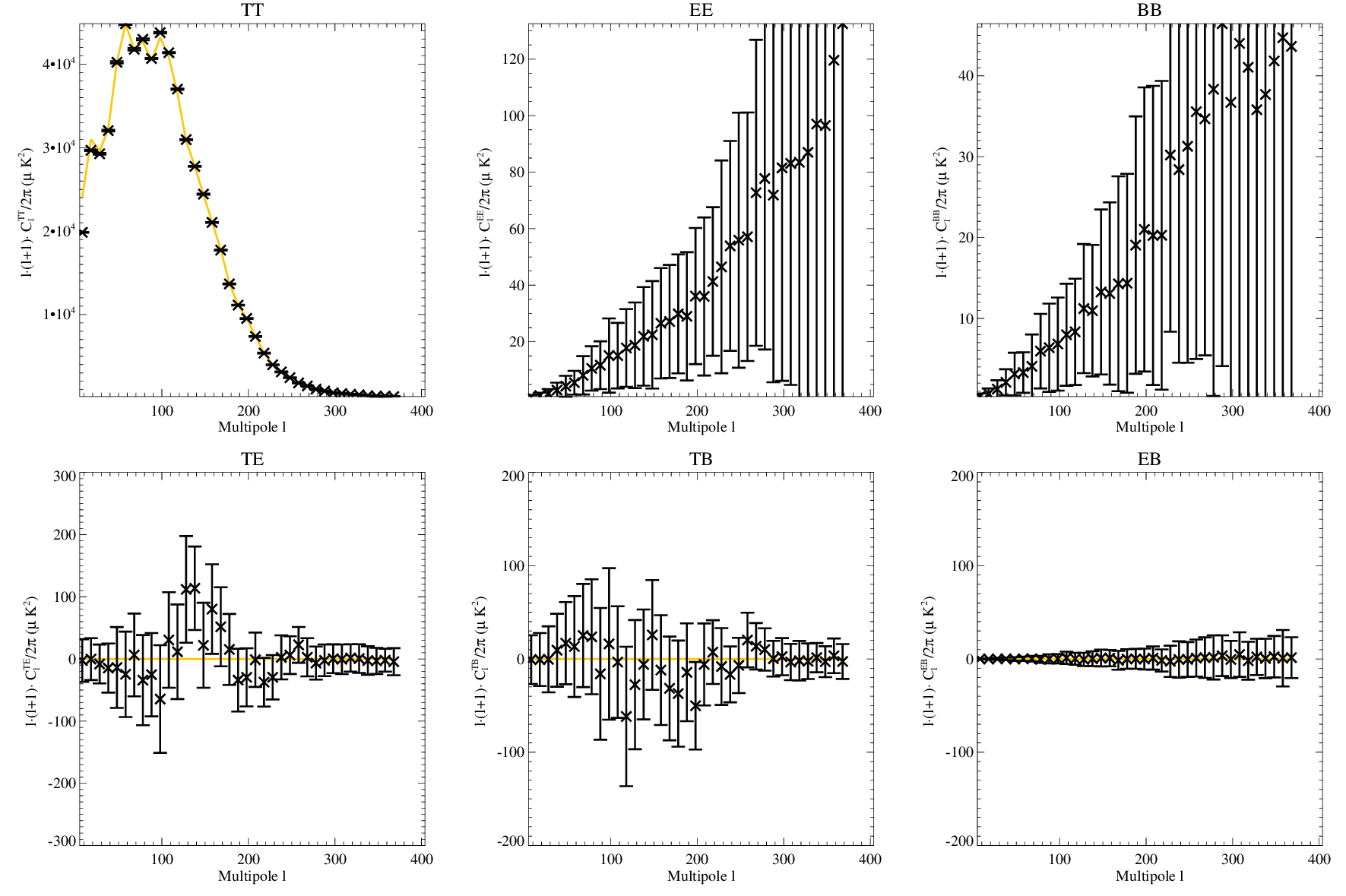}
\caption[fig:figure1]{\emph{A-fixed} reconstruction of the
  free-free emission $C_\ell^{TT}$, $C_\ell^{EE}$, $C_\ell^{BB}$, $C_\ell^{TE}$,
$C_\ell^{TB}$ and $C_\ell^{EB}$ power spectra (crosses),
 in $\mu {\rm K}_{\rm CMB}^2$, at 30
GHz in the case of the \emph{realistic} model with 4 components. We overplot
in orange the input model. Error bars are computed using
a total of 100 simulations. \label{4comp_realistic_ff}}
\end{figure*}

With this set of simulations, we have performed several types of
  separations. First, we have worked on a joint temperature and
  polarization analysis, similar to the one presented in section 
  \ref{testing}. Then we have considered temperature-only separation,
  and polarization-only separation. 
  For each of the described cases we have applied the algorithm with
  the different degrees of freedom presented in section
  \ref{testing}. 

  Models and recovered data are average over bins of size 10 in $\ell$
  beginning at $\ell=3$. Error bars presented in this section represent the
  dispersion over 100 simulations. Note
  that $n_{\rm side}=128$ allows a theoretical reconstruction up to
  $\ell=383$ in temperature and $\ell\sim256$ in polarization
  \cite{healpix}. CMB recovered spectra are plotted at 100 GHz while
  dust at 353 GHz and synchrotron and free-free at 30 GHz.

\subsection[]{Joint temperature and polarization separation}

 %The algorithm has been
%  applied considering 3 and 4 components.

%\subsubsection*{4-components separation}

We have first performed a joint temperature and polarization
  separation on the \emph{realistic} model. We present here the results obtained in
  the \emph{A-fixed} separation case, considering the 4 simulated components. In this particular case, the
  algorithm is very slow to converge and thus 40000 EM iterations have
  been run. The recovered angular power spectra for temperature and
  polarization are displayed from figure \ref{4comp_realistic_cmb} to
  figure \ref{4comp_realistic_ff}, respectively for CMB,
  \emph{realistic} dust, synchrotron and free-free emissions and
  compared to the input model.

  For the CMB component, on figure
  \ref{4comp_realistic_cmb} we can see that $C_\ell^{TT}$ and
  $C_\ell^{TE}$ are recovered accurately up to
  $\ell\sim400$. Recovered spectra for $C_\ell^{TB}$ and $C_\ell^{EB}$
  are compatible with zero as expected. $C_\ell^{EE}$ is recovered with
  a good accuracy up to $\ell\sim250$ and then is slightly biased due
  to pixelization problems in the HEALPix scheme. Finally,
  the $C_\ell^{BB}$ spectrum is recovered up to $\ell\sim90$ and then is
  biased with residual noise as discussed in section
  \ref{sec:colornoisemodel}.  

  The \emph{Realistic} dust component recovered spectra are displayed on
  figure \ref{4comp_realistic_dust}. We can see that $C_\ell^{TT}$,
  $C_\ell^{EE}$, $C_\ell^{BB}$, $C_\ell^{TE}$, $C_\ell^{TB}$ and
  $C_\ell^{TT}$ are recovered with a perfect accuracy up to
  $\ell=383$.  
  
  The recovered power spectra for the synchrotron component are
  displayed on figure \ref{4comp_realistic_sync}. Polarization only
  power spectra ($C_\ell^{EE}$, $C_\ell^{BB}$ and $C_\ell^{EB}$) are
  recovered in good agreement with the input
  model. $C_\ell^{TT}$ is well recovered up to $\ell\sim200$ but then
  converge sharply to a null signal and therefore a residual noise
  bias similar to the one of $C_\ell^{BB}$ occurs. This is not
  directly visible in the temperature spectrum but can be seen in the
  cross temperature and polarization spectra. Indeed, $C_\ell^{TE}$
  and $C_\ell^{TB}$ are well recovered up to $\ell\sim200$ and further
  many points are strongly biased.

  Finally, results corresponding to the free-free emission component
  are displayed in figure \ref{4comp_realistic_ff}. $C_\ell^{TT}$ is
  recovered with a good accuracy up to $\ell=383$. As for polarization
  the input signal is null, we can see in the recovered spectra the overall behavior of
  our algorithm described in section \ref{testing}.
% $C_\ell^{EE}$ and
%  $C_\ell^{BB}$ recovered spectra contain residual noise from the
%  separation while cross power spectra do not, because of the fact
%  that noise is assumed uncorrelated between $I$, $Q$ and $U$ maps.

  The algorithm has also been run for \emph{CMB-fixed} and
  \emph{Blind} separations. For both of them, excepting the dust 
  component which is well constrained in all cases due to its dominant
  power at high frequencies, the algorithm fails to converge and then
  components are mixed and results strongly biased. This may come from
  the fact that free-free and synchrotron electromagnetic spectra are similar and to the fact that all the Galactic
  emissions have strong spatial correlations. In the following
  section we will address this problem and show that this mixing that
  prevents the convergence of the algorithm is mainly due to the separability problem
  that occurs in temperature. In section \ref{pol_sep}, we will see
  that the separation performed on sets of $Q$ and $U$ maps has not
  this separability problem. 

  Notice that when not considering the free-free emission in the
  simulations, results are very similar and the same performances 
  of the algorithm with respect to the level of prior we assume are
  observed. 

%\subsubsection*{3-components separation}

%\begin{figure*}
%\centering
% \includegraphics[width=17cm,height=10cm]{./Pol_CMB_cmb_sync_dust_ff_128_3000_EM_CMBfixed_trueA_100simus_3comp.ps}
%\caption[fig:figure1]{\label{3comp_realistic_cmb}}
%\end{figure*}

%\begin{figure*}
%\centering
% \includegraphics[width=17cm,height=10cm]{./Pol_SYNC_cmb_sync_dust_ff_128_3000_EM_CMBfixed_trueA_100simus_3comp.ps}
%\caption[fig:figure1]{\label{3comp_realistic_sync}}
%\end{figure*}

\subsection[]{Temperature-only separation}
\label{tt_sep}

\begin{figure*}
\centering
 \includegraphics[width=17cm,height=5cm]{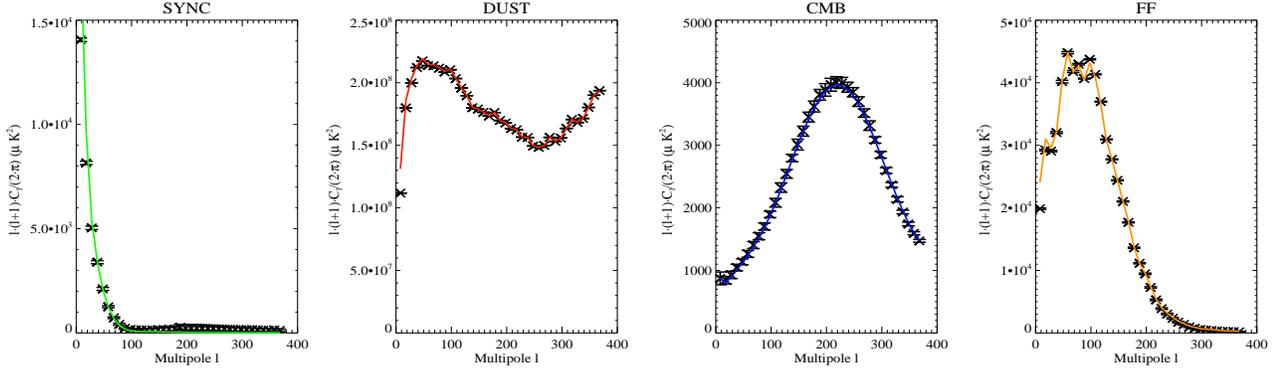}
\caption[fig:figure1]{\emph{A-fixed} temperature-only reconstruction of the
  $C_\ell^{TT}$ power spectra (crosses) in $\mu {\rm K}_{\rm CMB}^2$ in the
  case of the \emph{realistic} model with 4 components. Results are
  displayed from left to right for the 4 different components, i.e
  synchrotron (at 30 GHz), \emph{realistic} dust (at 353 GHz), CMB (at
  100 GHz) and free-free (at 30 GHz). Input models
  are overplotted in solid lines. Error bars are computed using
a total of 100 simulations. \label{tt_4comp}}
\end{figure*}

\begin{figure*}
\centering
 \includegraphics[width=17cm,height=5cm]{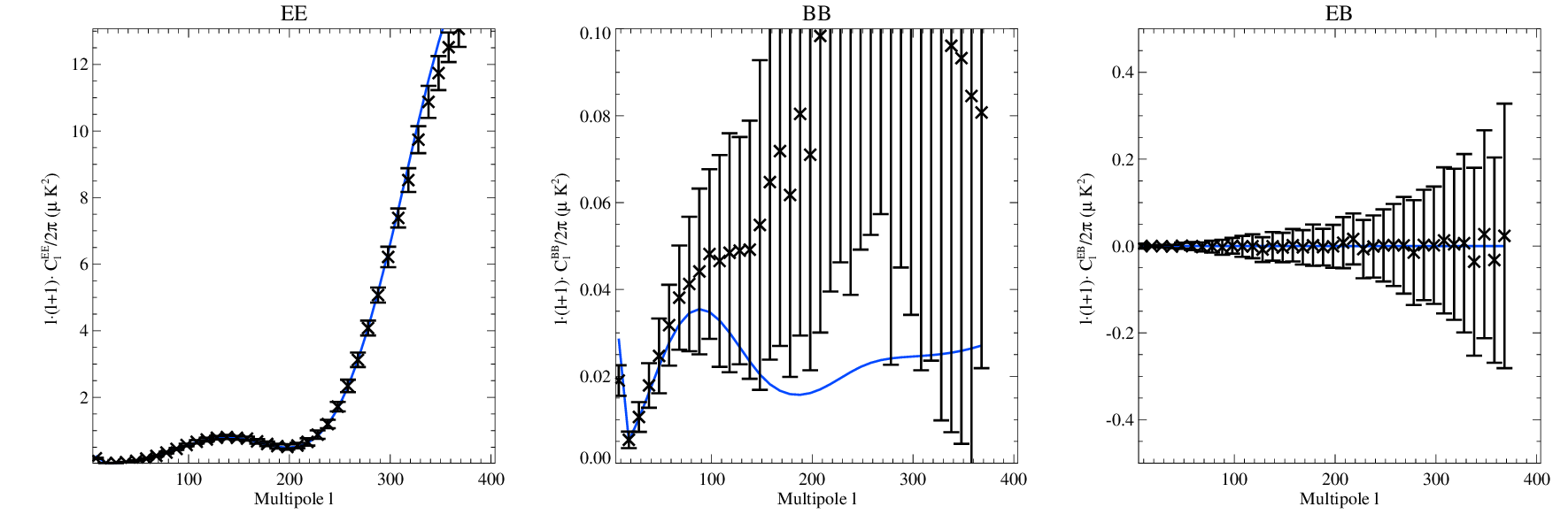}
\caption[fig:figure1]{\emph{A-fixed} polarization-only reconstruction
  of the CMB $C_\ell^{EE}$, $C_\ell^{BB}$ and $C_\ell^{EB}$
  power spectra (crosses) in $\mu {\rm K}_{\rm CMB}^2$ at 100 GHz in the
  case of the \emph{realistic} model with 3 polarized components. Input models
  are overplotted in blue. Error bars are computed using
a total of 100 simulations. \label{eb_3comp_cmb}}
\end{figure*}

\begin{figure*}
\centering
 \includegraphics[width=17cm,height=5cm]{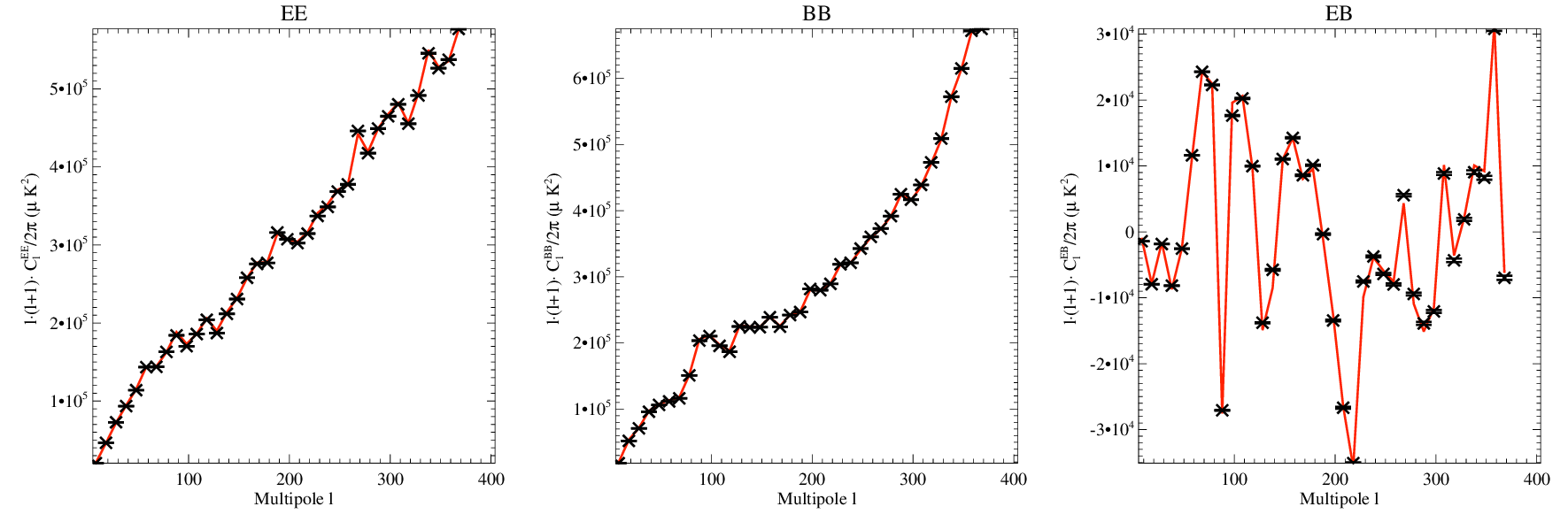}
\caption[fig:figure1]{\emph{A-fixed} polarization-only reconstruction
  of the \emph{realistic} dust $C_\ell^{EE}$, $C_\ell^{BB}$ and $C_\ell^{EB}$
  power spectra (crosses) in $\mu {\rm K}_{\rm CMB}^2$ at 353 GHz in the
  case of the \emph{realistic} model with 3 polarized components. Input models
  are overplotted in red. Error bars are computed using
a total of 100 simulations. \label{eb_3comp_dust}}
\end{figure*}

\begin{figure*}
\centering
 \includegraphics[width=17cm,height=5cm]{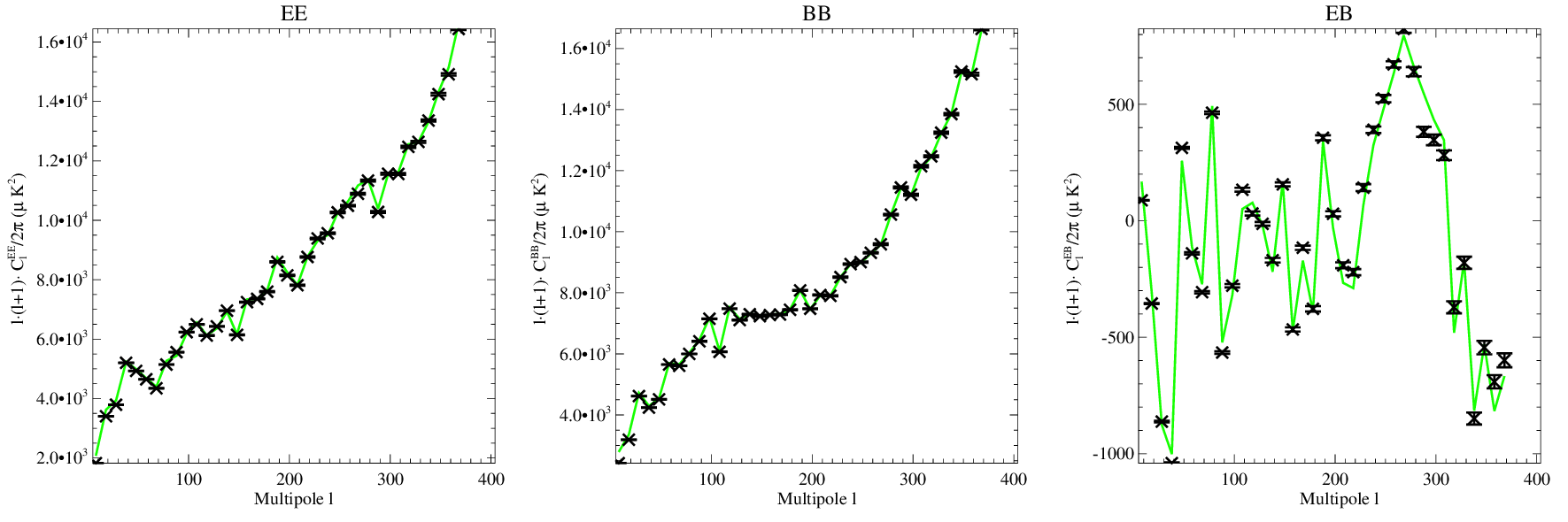}
\caption[fig:figure1]{\emph{A-fixed} polarization-only reconstruction
  of synchrotron emission $C_\ell^{EE}$, $C_\ell^{BB}$ and $C_\ell^{EB}$
  power spectra (crosses) in $\mu {\rm K}_{\rm CMB}^2$ at 30 GHz in the
  case of the \emph{realistic} model with 3 polarized components. Input models
  are overplotted in green. Error bars are computed using
a total of 100 simulations. \label{eb_3comp_sync}}
\end{figure*}

\begin{figure*}
\centering
 \includegraphics[width=17cm,height=5cm]{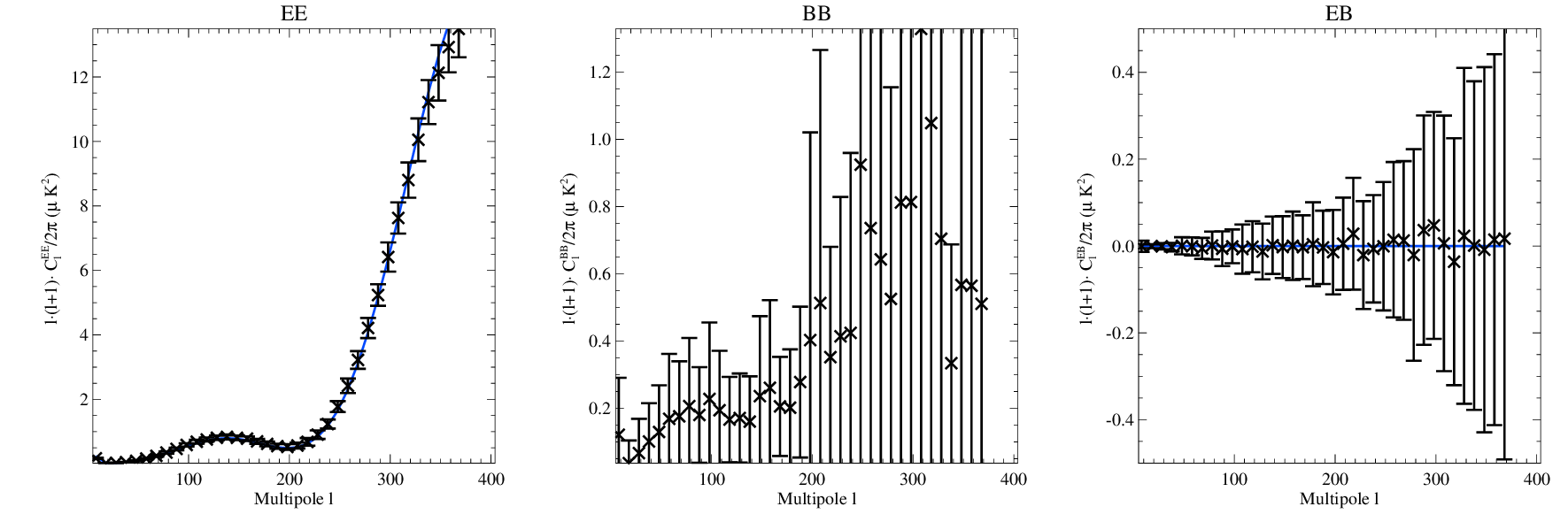}
\caption[fig:figure1]{\emph{Blind} polarization-only reconstruction
  of the CMB $C_\ell^{EE}$, $C_\ell^{BB}$ and $C_\ell^{EB}$
  power spectra (crosses) in $\mu {\rm K}_{\rm CMB}^2$ at 100 GHz in the
  case of the \emph{realistic} model with 3 polarized components. Input models
  are overplotted in blue. Error bars are computed using
a total of 100 simulations. \label{eb_3comp_cmb_blind}}
\end{figure*}

\begin{figure*}
\centering
 \includegraphics[width=17cm,height=5cm]{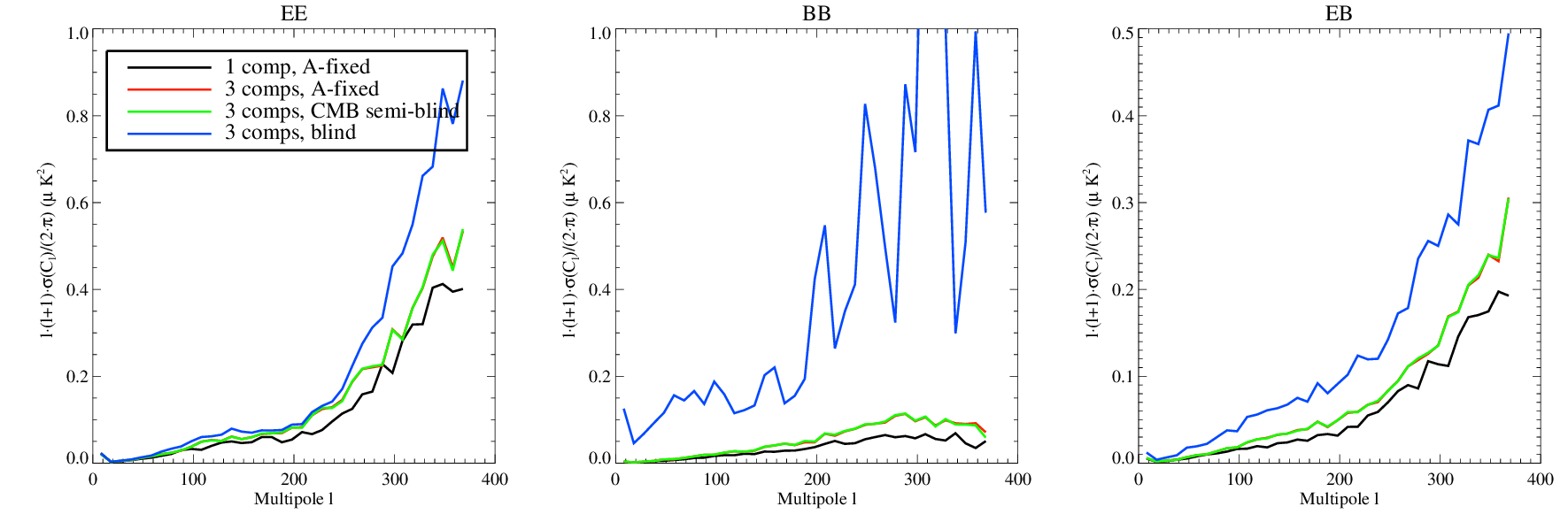}
\caption[fig:figure1]{Error bars of the reconstruction of the CMB
  $C_\ell^{EE}$, $C_\ell^{BB}$ and $C_\ell^{EB}$ power spectra in $\mu
  {\rm K}_{\rm CMB}^2$ at 100 GHz in a \emph{Blind} (blue), \emph{CMB
    semi-blind} (green) and \emph{A-fixed} separations of a 3
  components mixture containing CMB, \emph{realistic} dust and
  synchrotron and in a \emph{A-fixed} separation of a set CMB and
  noise only simulations (black). \label{compa_err_bars_eb}}
\end{figure*}

We have performed a temperature-only separation on the
  \emph{realistic} model. For this we consider sets of $I$ maps and
  the algorithm solve the spectral matching equations for $T$ modes
  only, like in the SMICA algorithm \cite{delabrouille}. We present here the results obtained in
  the \emph{A-fixed} separation case, considering the 4 simulated
  components.

  Recovered $C_\ell^{TT}$ spectra are displayed in figure
  \ref{tt_4comp} for synchrotron, \emph{realistic} dust, CMB and
  free-free. We can see that except the synchrotron spectrum which
  start to be biased at $\ell\sim200$, for the reason advanced in the
  last section, spectra are recovered with a good accuracy for the
  dust, CMB and free-free components up to $\ell=383$. We have also performed the separation
  for the \emph{CMB-fixed} and the \emph{Blind} cases. As for the
  joint analysis, excepting again the dust component for which the
  spectrum is recovered efficiently, the algorithm fails to
  converge. Note that in the literature, no method has successfully
  separated synchrotron from dust on a noisy
  simulated mixture of CMB and astrophysical foregrounds, working on
  all sky maps.

  Comparable results and the same performances
  of the algorithm with respect to the
  level of prior we assume are observed when no free-free component is
  included in simulations.

\subsection[]{Polarization-only separation}
\label{pol_sep}

Finally, we have performed a polarization-only separation on the
  \emph{realistic} model. For this we consider sets of $Q$ and $U$ maps and
  the algorithm solve the spectral matching equations for $E$ and $B$
  modes (allowing reconstruction of $C_\ell^{EE}$, $C_\ell^{BB}$ and
  $C_\ell^{EB}$). Notice that as we suppose that the free-free emission
  is not polarized, the sets of polarized maps used here only contain
  CMB, \emph{realistic} dust, synchrotron and noise.

\subsubsection*{\emph{CMB semi-blind} separation}

We have performed a \emph{CMB semi-blind} separation on the
  polarization-only set of maps. In this case, the CMB electromagnetic
  spectrum is initialized to 1 and kept fixed (see section~\ref{testing}).

  The results of this analysis on the CMB power spectra are displayed in figure
  \ref{eb_3comp_cmb}. We can see that $C_\ell^{EE}$ are
  reconstructed without bias up to $\ell\sim300$, as expected with
  this pixelization scheme and that $C_\ell^{EB}$ is compatible with
  zero. $C_\ell^{BB}$ is efficiently recovered up to $\ell\sim50$ and
  then is biased with residual noise from the separation.
% This is poor
%  compared to the joint temperature and polarization case presented
% above but we want to stress the fact that in the former the whole
%  $A$ matrix was fixed to its true value.

  The \emph{realistic} dust and synchrotron emissions recovered power spectra are displayed in
  figure \ref{eb_3comp_dust} and \ref{eb_3comp_sync}. In both cases we
  can see that $C_\ell^{EE}$, $C_\ell^{BB}$ and $C_\ell^{EB}$ are
  reconstructed very efficiently up to $\ell=383$.

%  These results shows that in the case of a polarization-only
%  analysis it is possible to have an efficient separation, without
%  having the exact mixing matrix as a starting point of the
%  algorithm. This is true when considering polarization only. The
%  strong spatial correlations that exists between Galactic components
%  in temperature prevents to do so. 

\subsubsection*{\emph{Blind} separation}

We have also performed, for the polarization-only separation, a
  \emph{Blind} separation. The results of this analysis for dust and synchrotron are not
  displayed, but for both, $C_\ell^{EE}$, $C_\ell^{BB}$ and $C_\ell^{EB}$ are
  reconstructed very efficiently up to $\ell=383$. The reconstruction of
  the CMB component is displayed in figure
  \ref{eb_3comp_cmb_blind}. For $C_\ell^{EE}$ we can see that the
  reconstruction is similar to the one in the \emph{CMB semi-blind}
  case, only the error bars are larger. However, the $C_\ell^{BB}$
  spectrum is not recovered.% This means that because of 
%  the faint power of the $B$ modes, the algorithm need to have at
%  least the CMB electromagnetic spectrum fixed in order to recover the
%  $C_\ell^{BB}$ signal.

\subsection[]{Discussions}
\label{realistic_discussions}

From the previous analyzes we have clearly identified a
  separability problem when dealing with more than one realistic diffuse
  Galactic emission component. This problem appears both in the joint
  temperature and polarization and in the temperature-only analyzes,
  but not in the polarization-only one. This would indicate that it
  is mainly due to the high level of correlation of the Galactic
  diffuse emission in temperature both in the Galactic plane and at
  high Galactic latitudes. The current version of our algorithm assume
  uncorrelated components and therefore we expect it to behave badly
  when they are correlated. Work is in progress to adapt {\sc
    PolEMICA} to account for spatially correlated components.

  For polarization-only separation the correlation problem seems to
  be not significant and the \emph{CMB semi-blind} and \emph{blind}
  analyzes are possible. Therefore, for this case, we can evaluate the
  loss of accuracy in the reconstruction of the CMB signal due to the
  foreground contamination. For this, we compare the error bars on the
  CMB power spectra for the \emph{A-fixed}, \emph{CMB semi-blind} and
  \emph{blind} separations to the one obtained in the case of a
  \emph{A-fixed} separation on {\bf \tt
    [planck~b]} simulations containing CMB and noise only.

%{\bf In summary of the analysis of our \emph{realistic} model, we have
%  seen that the separability problem that occurs comes from the
%  separation of the galactic component in temperature. If the
%  algorithm is launched without the real mixing matrix as a starting
%  point and if the mixing matrix elements are not fixed, a mixing
%  between Galactic components occur and prevents the algorithm to
%  converge. This remains true if we make a joint temperature and
%  polarization separation. Indeed, considering in the likelihood
%  function both the temperature and the polarization information, the
%  mixing that occurs in temperature prevents the whole algorithm to
%  succeed. Unless we fix the $A$ matrix to its input value. When
%  considering only polarization, a \emph{CMB semi-blind} separation
%  for both $E$ and $B$ modes and a \emph{Blind} separation for the $E$
%  modes become possible and allow a efficient reconstruction. It may
%  be possible to include the spatial correlation that exists in
%  temperature between Galactic components in the maximization of the
%  likelihood. Work is in progress for taking into account this
%  separability problem in our algorithm. 

  Results are
  presented in figure \ref{compa_err_bars_eb}. $C_\ell^{EE}$,
  $C_\ell^{BB}$ and $C_\ell^{EB}$ reconstruction error bars behave similarly with respect to the different algorithm
  priors. For each of them, \emph{A-fixed} and \emph{CMB semi-blind}
  error bars are of the same amplitude and are between 25 and 50 \% larger
  (respectively a factor 1.26 for $C_\ell^{EE}$, 1.50 for
  $C_\ell^{BB}$ and 1.29 for $C_\ell^{EB}$) than the reference
  \emph{A-fixed} CMB only case. This means that in the context of our
  \emph{realistic} model, we have no need
  to put priors on the foregrounds electromagnetic spectra in
  polarization to perform an efficient separation. In the \emph{Blind}
  case, error bars of the reconstruction are increased by a factor
  1.59 for $C_\ell^{EE}$, 13.5 for $C_\ell^{BB}$ (but the reconstruction
  is biased) and 2.52 for $C_\ell^{EB}$. On the
  other hand, these results stress the fact that having priors on the
  CMB electromagnetic spectra inside of our algorithm helps to perform
  a more accurate separation.

%+++++++++++++++++++++++++++++++++++++++++++++++++++++++++++++++++++++++++++++
%+++++++++++++++++++++++++++++++++++++++++++++++++++++++++++++++++++++++++++++
%+++++++++++++++++++++++++++++++++++++++++++++++++++++++++++++++++++++++++++++
%+++++++++++++++++++++++++++++++++++++++++++++++++++++++++++++++++++++++++++++
%+++++++++++++++++++++++++++++++++++++++++++++++++++++++++++++++++++++++++++++
%+++++++++++++++++++++++++++++++++++++++++++++++++++++++++++++++++++++++++++++

%+++++++++++++++++++++++++++++++++++++++++++++++++++++++++++++++++++++++++++++
%+++++++++++++++++++++++++++++++++++++++++++++++++++++++++++++++++++++++++++++
%+++++++++++++++++++++++++++++++++++++++++++++++++++++++++++++++++++++++++++++
%+++++++++++++++++++++++++++++++++++++++++++++++++++++++++++++++++++++++++++++
%+++++++++++++++++++++++++++++++++++++++++++++++++++++++++++++++++++++++++++++
%+++++++++++++++++++++++++++++++++++++++++++++++++++++++++++++++++++++++++++++

\section[]{Summary and conclusions}

%+++++++++++++++++++++++++++++++++++++++++++++++++++++++++++++++++++++++++++++
%+++++++++++++++++++++++++++++++++++++++++++++++++++++++++++++++++++++++++++++
%+++++++++++++++++++++++++++++++++++++++++++++++++++++++++++++++++++++++++++++
%+++++++++++++++++++++++++++++++++++++++++++++++++++++++++++++++++++++++++++++
%+++++++++++++++++++++++++++++++++++++++++++++++++++++++++++++++++++++++++++++
%+++++++++++++++++++++++++++++++++++++++++++++++++++++++++++++++++++++++++++++

\label{conclusions}
We present in this paper the {\sc PolEMICA} algorithm which is an extension to polarization
of the SMICA temperature MD-MC blind component separation method developed by \cite{delabrouille}.
Both algorithms work in harmonic space and are based on the spectral matching of the data to a noisy 
linear mixture of uncorrelated physical components using the EM algorithm to maximize the likelihood function.
By contrast to the temperature data which are described by a single scalar quantity $T$, 
the combined temperature and polarization data are described in harmonic space by three correlated
scalar quantities $T$, $E$ and $B$ corresponding to the $I$, $Q$ and $U$ Stokes parameters in real space.
We have developed a new formalism to jointly deal with the 6 resulting
auto and cross angular power spectra, $TT$, $EE$, $BB$, $TE$, $TB$ and $EB$. 

Using this formalism we have constructed the likelihood function and 
proved that the EM algorithm can be also applied to polarization data.
Under the assumption of uncorrelated Gaussian distributed components and noise,
the free parameters in the fit are the mixing matrix describing the
electromagnetic spectrum of the physical components for $T$, $E$ and $B$, 
the temperature and polarization angular power spectra of the physical components 
and the temperature and
polarization noise power spectra for each of the detectors.  \\

We have, as a first approach, intensively and successfully tested the {\sc PolEMICA} method on simulations of the Planck satellite experiment
considering a 14-months nominal mission and no systematic effects.
For these tests, we suppose a \emph{simplified} linear model for the sky emission including CMB, synchrotron
with constant spectral index and \emph{simplified}-dust (Gaussian realization) emissions. We construct full sky maps 
for all the  polarized channels from 30 to 353~GHz including at least one of the above
physical components and considering white noise and infinite resolution. 

The method permits blind separation on these simulations allowing us to reconstruct
the noise and physical component's temperature and polarization power spectra 
as well as the mixing matrix when we consider equal electromagnetic spectrum in $T$, $E$
and $B$. When we relax this hypothesis the reconstruction of the electromagnetic
spectrum for the CMB $B$ modes is significantly degraded as could be expected
because of the low signal to noise ratio. These results indicate that the
{\sc PolEMICA} method allows us to both constrain the electromagnetic spectrum of the
physical components and also to inter-calibrate the data based on the
reconstructed CMB electromagnetic spectrum. \\

After setting the general performances of the algorithm, we have performed the separation on a more
  \emph{realistic} model that includes \emph{realistic}-dust,
  synchrotron and
  free-free components in section \ref{realistic_model}. We have encountered in this case a
  separability problem, that mixes up components and prevents the
  algorithm to converge, when performing \emph{blind} separations. We
  have shown that this is due to spatial correlations between Galactic
  components in temperature. Thus, when working on sets of $Q$ and $U$
  maps and maximizing the likelihood for $EE$, $BB$ and $EB$ modes
  only, this separability problem does not occur and \emph{CMB
    semi-blind} and \emph{blind} separations are possible.
  For this \emph{polarization-only} case, we have shown that
  considering our \emph{realistic} sky model and our algorithm, in the Planck case, we
  have no need to put priors on the Galactic components
  electromagnetic spectra to reconstruct the CMB polarized power spectra. Nevertheless, adding priors on the CMB
  electromagnetic spectrum helps to perform a more accurate separation.

%To evaluate the impact of foregrounds in the determination of the CMB temperature and polarization
%power spectra we have compared the results of the analysis on CMB only and CMB plus 
%foregrounds simulations.
%In the presence of foregrounds, the error bars on the reconstruction of the CMB power spectra 
%are increased by at least a factor of two both in temperature and in polarization
%even if we consider a {\it CMB semi-blind} analysis assuming the CMB electromagnetic spectrum known.
%Therefore, although the foreground contribution in the data can be removed,
%it significantly reduces the precision to which the CMB polarization
%signal can be extracted from the data. 
%For a {\it blind} analysis the error bars are larger 
%indicating that the use of prior information is badly needed
%for the analysis of future CMB data sets. This work is
%based on a simplified foreground model. Thus, quantitative results
%presented here are not directly applicable to real data but are just
%good qualitative indications of what may happen on real data.\\

Finally, real experiments present finite resolution,
partial effective sky coverage, systematic effects and, often, correlated noise.
All these issues must be dealt with by the component separation algorithms and 
will with no doubt significantly limit the precision to which the CMB signal may be reconstructed.  
{\sc PolEMICA}, as it was already the case for SMICA, can account for beam 
and filtering smoothing. Systematic effects and correlated noise can be modeled as extra
components in the data for which the spectral dependence can be estimated in
a {\it blind} analysis.
Moreover, the strong spatial correlation in temperature between Galactic physical emissions:
  dust, synchrotron and free-free, is a major problem for 
  blind component separation algorithms which generally assumed uncorrelated components.
  Although not observed yet, we can also imagine spatial correlation of the
  Galactic emissions in polarization. Work is in progress to adapt the {\sc PolEMICA}
  algorithm to the case of correlated components.
 
%  VOICI TA PHRASE
%  Work is in progress to make this algorithm work on a more
%  realistic sky model containing dust emission that do vary with the
%  Galactic latitude. We still have problems to converge in temprature
%  while the separation is efficient when dealing with polarization alone}.
In addition, foreground emissions have in general spatially varying 
electromagnetic spectra far beyond the simple linear model presented here. 
Work is also in progress to adapt {\sc PolEMICA} to the case of foregrounds
with spatially varying electromagnetic spectrum.

%________________________________________________________________
\section*{Acknowledgments}
We would like to thank D. Santos and F.X~D\'esert for very
useful comments and a careful reading of the paper.
Special thanks to M. Tristram for his comments and power spectrum
related procedures and to D. Blais for his useful advises on matrix derivation.
We acknowledge J.F. Cardoso, J. Delabrouille and 
G. Patanchon for comments on the technical details of the algorithm.
The HEALPix package \cite{healpix} was  used extensively in this paper.
%\dots\dots\dots
%  Authors are grateful to E. Battaner for discussions in the
%  initial stage of this work in our team. We also want to thank
%  F.X~D\'esert for useful discussions on the method and for
%  corrections of this paper. 

%________________________________________________________________

%________________________________________________________________
%\onecolumn

%===========================================================================

\onecolumn
\newpage
\appendix
\section[]{MD-MC Polarization sky model}\label{matrix_appendix}

\noindent We discuss here the formalism developed to describe the temperature
and polarization observations as a noisy mixture of independent components.

\noindent In the following we assume full sky observations at two frequencies
$\nu_1$ and $\nu_2$ and a simple linear model for the sky emission
with two components  $c_1$ and $c_2$.

\noindent In this case, equation (\ref{data}) reads

$$
\left(\begin{array}{c}
y^{\nu_1T}_{\ell m}\\
y^{\nu_1E}_{\ell m}\\
y^{\nu_1B}_{\ell m}\\
y^{\nu_2T}_{\ell m}\\
y^{\nu_2E}_{\ell m}\\
y^{\nu_2B}_{\ell m}\end{array}\right)= \left(\begin{array}{cccccc}
f^T_{\nu_1c_1}&0&0&f^T_{\nu_1c_2}&0&0\\
0&f^E_{\nu_1c_1}&0&0&f^E_{\nu_1c_2}&0\\
0&0&f^B_{\nu_1c_1}&0&0&f^B_{\nu_1c_2}\\
f^T_{\nu_2c_1}&0&0&f^T_{\nu_2c_2}&0&0\\
0&f^E_{\nu_2c_1}&0&0&f^E_{\nu_2c_2}&0\\
0&0&f^B_{\nu_2c_1}&0&0&f^B_{\nu_2c_2}\end{array}\right)\cdot
\left(\begin{array}{c}
s^{c_1T}_{\ell m}\\
s^{c_1E}_{\ell m}\\
s^{c_1B}_{\ell m}\\
s^{c_2T}_{\ell m}\\
s^{c_2E}_{\ell m}\\
s^{c_2B}_{\ell m}\end{array}\right)+ \left(\begin{array}{c}
n^{\nu_1T}_{\ell m}\\
n^{\nu_1E}_{\ell m}\\
n^{\nu_1B}_{\ell m}\\
n^{\nu_2T}_{\ell m}\\
n^{\nu_2E}_{\ell m}\\
n^{\nu_2B}_{\ell m}\end{array}\right)
$$

\noindent where $y^{\nu_iX}_{\ell m}$ and $s^{c_jX}_{\ell m}$ for $X= \lbrace T, E, B \rbrace$ 
are the coefficients of the spherical harmonic decomposition of the 
input sky observations and of the components of the sky model respectively.
The coefficients $f^X_{\nu_ic_j}$ correspond to 
the electromagnetic spectrum of the component $c_j$ at the frequency
of observation $\nu_{i}$. Note that the mixing matrix, $A$, has dimensions 
$3\cdot n_\nu\times3\cdot n_c$. \\

\noindent We define the noise, $R_n(b)$, signal, $R_s(b)$, and data, $R_y(b)$, density matrices used 
in equation (\ref{dataspec}) for each bin, $b$, as follows

$$
R_z(b)\equiv\frac{1}{n_b}\sum_{\vec{\ell}\in\mathcal{D}_b}<z(\vec{\ell})z(\vec{\ell})^\dagger>\
,\qquad z=\{n,s,y\}
$$

\indent Assuming the noise uncorrelated between detectors the noise density matrix is
diagonal 

$$R_n(b)=\left(\begin{array}{ccc}
    &\nu_1&\nu_2\\
\nu_1 &
    \begin{array}{ccc}
    N_\ell^{TT}(b)&&\\
    &N_\ell^{EE}(b)&\\
    &&N_\ell^{BB}(b)\\
    \end{array}
&
    \begin{array}{ccc}
    &&\\
    &0&\\
    &&\\
    \end{array}
\\
\nu_2 &
    \begin{array}{ccc}
    &&\\
    &0&\\
    &&\\
    \end{array}
&
    \begin{array}{ccc}
    N_\ell^{TT}(b)&&\\
    &N_\ell^{EE}(b)&\\
    &&N_\ell^{BB}(b)\\
    \end{array}
\\
\end{array}\right)
$$

\noindent In the same way assuming independent physical components the
signal density matrix is block diagonal and reads

$$R_s(b)=\left(\begin{array}{ccc}
    &c_1&c_2\\
c_1 &
    \left[\begin{array}{ccc}
    S_\ell^{TT}(b)&S_\ell^{TE}(b)&S_\ell^{TB}(b)\\
    S_\ell^{TE}(b)&S_\ell^{EE}(b)&S_\ell^{EB}(b)\\
    S_\ell^{TB}(b)&S_\ell^{EB}(b)&S_\ell^{BB}(b)\\
    \end{array}\right]
&
    \begin{array}{ccc}
    &&\\
    &0&\\
    &&\\
    \end{array}
\\
c_2 &
    \begin{array}{ccc}
    &&\\
    &0&\\
    &&\\
    \end{array}
&
    \left[\begin{array}{ccc}
    S_\ell^{TT}(b)&S_\ell^{TE}(b)&S_\ell^{TB}(b)\\
    S_\ell^{TE}(b)&S_\ell^{EE}(b)&S_\ell^{EB}(b)\\
    S_\ell^{TB}(b)&S_\ell^{EB}(b)&S_\ell^{BB}(b)\\
    \end{array}\right]
\\
\end{array}\right)$$

\noindent Finally the data density matrix can written by  blocks as
follows

$$R_y(b)=
\left(\begin{array}{ccc}
    &\nu_1&\nu_2\\
\nu_1 &
    \left[\begin{array}{ccc}
    Y_\ell^{TT}(b)&Y_\ell^{TE}(b)&Y_\ell^{TB}(b)\\
    Y_\ell^{TE}(b)&Y_\ell^{EE}(b)&Y_\ell^{EB}(b)\\
    Y_\ell^{TB}(b)&Y_\ell^{EB}(b)&Y_\ell^{BB}(b)\\
    \end{array}\right]
&
    \left[\begin{array}{ccc}
    Y_\ell^{TT}(b)&Y_\ell^{TE}(b)&Y_\ell^{TB}(b)\\
    Y_\ell^{TE}(b)&Y_\ell^{EE}(b)&Y_\ell^{EB}(b)\\
    Y_\ell^{TB}(b)&Y_\ell^{EB}(b)&Y_\ell^{BB}(b)\\
    \end{array}\right]
\\
\nu_2 &
    \left[\begin{array}{ccc}
    Y_\ell^{TT}(b)&Y_\ell^{TE}(b)&Y_\ell^{TB}(b)\\
    Y_\ell^{TE}(b)&Y_\ell^{EE}(b)&Y_\ell^{EB}(b)\\
    Y_\ell^{TB}(b)&Y_\ell^{EB}(b)&Y_\ell^{BB}(b)\\
    \end{array}\right]
&
    \left[\begin{array}{ccc}
    Y_\ell^{TT}(b)&Y_\ell^{TE}(b)&Y_\ell^{TB}(b)\\
    Y_\ell^{TE}(b)&Y_\ell^{EE}(b)&Y_\ell^{EB}(b)\\
    Y_\ell^{TB}(b)&Y_\ell^{EB}(b)&Y_\ell^{BB}(b)\\
    \end{array}\right]
\\
\end{array}\right)$$

\end{document}